\documentclass[letterpaper,journal,onecolumn]{IEEEtran}

\pdfoutput=1

\usepackage{amsfonts}
\usepackage{amssymb}
\usepackage{upgreek}
\usepackage{setspace}
\usepackage{multicol}
\usepackage{cite}
\usepackage{ctable}
\usepackage[cmex10]{amsmath}
\usepackage{array}
\usepackage{mdwmath}
\usepackage{mdwtab}
\usepackage{eqparbox}
\usepackage{url}
\usepackage{graphicx}
\usepackage{amsmath}
\usepackage{epsf}
\usepackage{mathrsfs}
\usepackage{cite}
\usepackage{dsfont}
\usepackage[usenames,dvipsnames]{pstricks}
\usepackage{epsfig}
\usepackage{enumerate}
\usepackage{hyperref}
\usepackage{cleveref}
\usepackage{mdwlist}

\newtheorem{theorem}{Theorem}
\newtheorem{lemma}{Lemma}
\newtheorem{remark}{Remark}

\newcommand{\xu}{\mathbf{u}}
\newcommand{\xv}{\mathbf{v}}

\newcommand{\xy}{\mathbf{y}}
\newcommand{\xa}{\mathbf{a}}
\newcommand{\xb}{\mathbf{b}}
\newcommand{\xc}{\mathbf{c}}
\newcommand{\xd}{\mathbf{d}}
\newcommand{\xn}{\mathbf{n}}
\newcommand{\xh}{\mathbf{h}}
\newcommand{\xG}{\mathbf{G}}
\newcommand{\xH}{\mathbf{H}}

\newcommand{\xP}{\mathbf{P}}
\newcommand{\xQ}{\mathbf{Q}}
\newcommand{\xR}{\mathbf{R}}
\newcommand{\xA}{\mathbf{A}}
\newcommand{\xB}{\mathbf{B}}
\newcommand{\xC}{\mathbf{C}}
\newcommand{\xD}{\mathbf{D}}
\newcommand{\xN}{\mathbf{N}}
\newcommand{\xPhi}{\mathbf{\Phi}}
\newcommand{\xOmega}{\mathbf{\Omega}}
\newcommand{\xomega}{\boldsymbol{\omega}}

\newcommand{\xxC}{\mathcal{C}}

\newcommand{\xxN}{\mathcal{N}}

\newcommand{\xxS}{\mathcal{S}}

\newcommand{\xxW}{\mathcal{W}}

\allowdisplaybreaks

\newcommand{\vect}[1]{\mathbf{#1}}
\newcommand{\Def}{\triangleq}

\newcommand{\DoFa}{\underline{\mathsf{DoF}}}
\newcommand{\DoF}{\mathsf{DoF}}
\newcommand{\ie}{i.e., }

\DeclareMathOperator{\rank}{rank}

\crefname{lemma}{Lemma}{Lemmas}
\Crefname{lemma}{Lemma}{Lemmas}
\crefname{theorem}{Theorem}{Theorems}
\Crefname{theorem}{Theorem}{Theorems}
\crefname{section}{Section}{Sections}
\Crefname{section}{Section}{Sections}
\crefname{figure}{Fig.}{Figs.}
\Crefname{figure}{Figure}{Figures}
\crefname{table}{Table}{Tables}
\Crefname{table}{Table}{Tables}
\crefname{equation}{\hspace{-1mm}}{Eqs.}
\Crefname{equation}{Equation}{Equations}

\makeatletter
\def\blfootnote{\xdef\@thefnmark{}\@footnotetext}
\makeatother

\begin{document}

\title{On the Degrees of Freedom of $K$-User SISO Interference and X Channels with Delayed CSIT}
\author{Mohammad~Javad~Abdoli,
Akbar~Ghasemi, and~Amir~Keyvan~Khandani
\thanks{This work was supported by Natural Sciences and Engineering Research Council of Canada (NSERC) and Ontario Ministry of Research and Innovation (ORF-RE). The material in this paper was presented in part at the 49th Annual Allerton Conference on Communication, Control, and Computing, Monticello, IL, September 2011.}%
\thanks{M. J. Abdoli is with Huawei Technologies Canada Co., Ltd., Kanata, ON K2K 3J1, Canada (e-mail: mjabdoli@uwaterloo.ca).}
\thanks{A. Ghasemi is with Ciena Corp., Ottawa, ON K2H 8E9, Canada (e-mail: aghasemi@uwaterloo.ca).}
\thanks{A. K. Khandani is with the Coding and Signal Transmission Laboratory, Department of Electrical and Computer Engineering, University of Waterloo, Waterloo, ON N2L 3G1, Canada (e-mail: khandani@uwaterloo.ca).}
\thanks{Copyright (c) 2013 IEEE. Personal use of this material is permitted. However, permission to use this material for any other purposes must be obtained from the IEEE by sending a request to pubs-permissions@ieee.org}
}

\maketitle

\begin{abstract}
The $K$-user single-input single-output (SISO)  AWGN interference channel and $2\times K$ SISO AWGN X channel are considered where the transmitters have the delayed channel state information (CSI) through noiseless feedback links. Multi-phase transmission schemes are proposed for both channels which possess novel ingredients, namely, multi-phase partial interference nulling, distributed interference management via user scheduling, and distributed higher-order symbol generation. The achieved degrees of freedom (DoF) values are greater than the best previously known DoFs for both channels with delayed CSI at transmitters.
\end{abstract}

\section{Introduction}
\label{Sec:Introduction}
The impact of feedback on the performance of a communication system has been extensively investigated from different perspectives. Although a negative result was first established by Shannon \cite{shannon1956zero} indicating that feedback does not increase the capacity of a memoryless point-to-point channel, there are various results affirming the significant effect of feedback on other performance criteria such as complexity and error probability of this channel.

In multi-terminal networks, however, it has been shown that feedback can enlarge the capacity region. The capacity region enlargements for multiple access and broadcast channels with access to noiseless output feedback are reported for additive white Gaussian noise (AWGN) cases in \cite{ozarow1984MACcapacity,ozarow1984BCCachievable} and for discrete cases in \cite{cover1981achievable,shayevitz2013capacity}. The capacity region enlargements in the AWGN multiple access and broadcast channels are ``additive'' which are bounded with increase of signal-to-noise ratio (SNR). The capacity region of a two-user AWGN interference channel under the assumption that each transmitter has access to a noiseless feedback link from its respective receiver is addressed in \cite{cadambe2008feedback,suh2011feedback}. It was shown that such feedback links can provide ``multiplicative'' gain in the capacity region of AWGN interference channels, i.e., the gap between the feedback and non-feedback capacity regions can be arbitrarily enlarged as SNR increases.

An important performance measure for a power constrained communication channel is its degrees of freedom (DoF) which determines behavior of the sum-capacity in high SNR regime. While it is a common assumption that receiver(s) know the channel state information (CSI) perfectly and instantaneously, the CSI knowledge of transmitter(s) (CSIT) is usually subject to some limitations. At one extreme, it is assumed that the transmitter(s) know the CSI instantaneously and perfectly (full CSIT assumption). Under this condition, the capacity region (and hence, the DoF region) of the multiple-input multiple-output (MIMO) broadcast channel was characterized in \cite{WeingarternMIMO_BC}. The DoF of the $K$-user single-input single-output (SISO) interference channel was shown to be $\frac{K}{2}$ with full CSIT \cite{cadambe2008interference}. It was shown in  \cite{cadambe2009X} that the $M\times K$ SISO X channel with full CSIT has $\frac{MK}{M+K-1}$ DoF. Also, in \cite{cadambe2009degrees}, it was proved that output feedback does not provide any DoF benefit in the interference and X channels under the full CSIT assumption. At the other extreme, the transmitter(s) are assumed to have no knowledge about CSI (no CSIT assumption). In this case, the $K$-user multiple-input single-output (MISO) broadcast channel was studied in \cite{Jafar2005IsotropicBC}. Other works include \cite{vaze2012NoCSIT} which characterizes the DoF regions of $K$-user MIMO broadcast, interference and X channels and \cite{huang2012degrees,zhu2012degrees,vaze2012new} which investigate the DoF region of two-user MIMO broadcast and interference channels with no CSIT. Specifically, by developing some upper and lower bounds, it was shown in \cite{vaze2012NoCSIT} that the MISO broadcast, SISO interference and SISO X channel under isotropic i.i.d. fading can achieve no more than one DoF.

The access of transmitter(s) to the full instantaneous CSI is an ideal assumption. While this assumption can be realized when the channel is subject to slow fading, it is not practically viable when channel variations are fast. In this situation, it is curious whether any type of feedback can improve the DoF performance of the channel. Recently, Maddah-Ali et al.\ in \cite{maddah2012completely} considered the MISO broadcast channel in fast fading environment where the transmitter obtains the past CSI through noiseless feedback (delayed CSIT). Surprisingly, they showed that even in an i.i.d. fading setup, the delayed version of CSIT significantly improves the channel DoF. In particular, it was shown that the $K$-user MISO broadcast channel with $M\geq K$ antennas at transmitter has $\frac{K}{1+\frac{1}{2}+\cdots+\frac{1}{K}}$ DoF under delayed CSIT assumption. Subsequently, the DoF region of the two-user MIMO broadcast channel with delayed CSIT was obtained in \cite{vaze2011DoF_BCC_Delayed}. Abdoli et al.\ in \cite{abdoli2011BCC} investigated the DoF of the symmetric three-user MIMO broadcast channel under delayed CSIT assumption. Achievable DoF regions for the two-user MIMO interference channel with delayed CSIT were obtained in \cite{Vaze2012IC_DCSIT,ghasemi2011interference}, and furthermore, by obtaining matching outer bounds, \cite{Vaze2012IC_DCSIT} characterized the exact DoF region. Maleki et al.\ in \cite{maleki2012retrospective} were the first to show that SISO channels with distributed transmitters can achieve more than one DoF under delayed CSIT assumption. In particular, they showed that the $2\times2$ SISO X and $3$-user SISO interference channel with delayed CSIT can achieve $\frac{8}{7}$ and $\frac{9}{8}$ DoF, respectively. Soon thereafter, Ghasemi et al.\ in \cite{Ghasemi2011Xchannel} achieved greater DoF of $\frac{6}{5}$ for the $2\times2$ X channel. They also proposed an immediate generalization of their transmission scheme to the $K\times K$ case, $K\geq2$, and achieved $\frac{4}{3}-\frac{2}{3(3K-1)}$ DoF with delayed CSIT. The DoF of the $2\times 2$ MIMO X channel with delayed CSIT was investigated for the symmetric case in \cite{ghasemi2012degrees}.

In this paper, whose results were partially reported in \cite{Abdoli2011Allerton}, we first investigate the problem of transmission over $K$-user SISO interference channel with delayed CSIT and $K\geq 3$. By proposing novel transmission schemes, we obtain achievable DoF results for this channel which are greater than the best previously known DoFs. Specifically, for the $3$-user case, we show that $\frac{36}{31}$ DoF is achievable which is greater than the previously reported $\frac{9}{8}$ DoF in \cite{maleki2012retrospective}. For the $K$-user case, $K>3$, we propose a multi-phase transmission scheme that achieves DoF values which are strictly greater than $\frac{36}{31}$ and approach the limiting value of $\frac{4}{6\ln 2 -1}\approx 1.2663$ as $K\to \infty$. 

Next, we consider the $2\times K$ SISO X channel with delayed CSIT and $K\geq 2$. Although we achieve the same $\frac{6}{5}$ DoF for the $2\times 2$ X channel as in \cite{Ghasemi2011Xchannel}, by proposing a multi-phase transmission scheme, we obtain achievable DoF values which are strictly greater than those obtained in \cite{Ghasemi2011Xchannel} for the $K\times K$ X channel with delayed CSIT and $K\geq3$. Our achievable DoFs for the $2\times K$ X channel approach the limiting value of $\frac{1}{\ln 2}\approx 1.4427$ as $K\to \infty$. We note that any achievable DoF in the $2\times K$ X channel is also achievable in the $K\times K$ X channel. 

We do not make any optimality claim on our achievable DoFs and the problem of DoF characterization of both channels with delayed CSIT remains open. However, if the channel coefficients are i.i.d.\ over users and time, we conjecture that the DoF of both the $K$-user interference channel and $M\times K$ X channel with delayed CSIT is bounded above by a constant, \ie it does not scale with number of users. In \cref{Sec:Conjecture}, we will provide some insights on this conjecture. 

The rest of the paper is organized as follows: the system model is described in \cref{Sec:SystemModel}. In \cref{Sec:PriorArt}, we summarize the related works by highlighting their main ideas and contributions. \Cref{Sec:MainResultsDiscussions} presents our main contributions and results. In \cref{Sec:SISO-IC,Sec:SISO-X}, our transmission schemes for SISO interference and X channels with delayed CSIT are elaborated on, respectively, and their achievable DoFs are obtained. Finally, \cref{Sec:Conclusion} concludes the paper.

\section{System Model}
\label{Sec:SystemModel}
For any integer $K\geq 2$, the discrete-time $K$-user SISO AWGN interference channel (IC) with private messages is defined by a set of $K$ transmitter-receiver pairs (TX$_i$,RX$_i$), ${1\leq i \leq K}$, where TX$_i$ wishes to communicate a message $W^{[i]}\in \xxW^{[i]}$ to RX$_i$. Moreover, the input-output relationship of this channel in time slot $t$, $t=1,2,\cdots$, is specified by
\begin{equation}
\label{Eq:IC-InputOutput}
y_j (t)= \sum_{i=1}^K h_{ji}(t) x_i(t) + z_j(t), \hspace{5mm} 1\leq j \leq K,
\end{equation}
where ${x_i(t)\in \mathbb{C}}$ is the complex channel input symbol transmitted by TX$_i$, ${y_j(t)\in \mathbb{C}}$ and ${z_j(t)\sim \xxC \xxN (0,1)}$ are the complex received symbol and additive complex Gaussian noise at the input of RX$_j$, respectively, and $h_{ji}(t)\in \mathbb{C}$ is the complex channel coefficient from TX$_i$ to RX$_j$. The channel input is subject to the average power constraint 
\begin{equation}
\label{Eq:PowerConstraint}
\frac{1}{\tau}\sum_{t=1}^\tau \mathbb{E}[|x_i(t)|^2]\leq P,
\end{equation}
where $\tau$ is the block length. The channel coefficients are assumed to be i.i.d. across the transmitters and receivers and to be drawn according to a continuous distribution, and the noise is assumed to be i.i.d. across receivers as well as time.

We further assume that the channel is subject to fast fading, \ie the channel coefficients are i.i.d. across time. Define the $K\times K$ channel matrix $\xH(t)\Def \left[ h_{ji}(t)\right]_{1\leq i,j\leq K}$. We make the following assumptions on knowledge of transmitters and receivers about the CSI:
\begin{itemize}
\item At the beginning of time slot $t$, $t\geq 2$, all transmitters perfectly know $\left\{ \xH(t')\right \}_{t'=1}^{t-1}$ via feedback links.
\item By the end of transmission block, \ie $t=\tau$, all receivers perfectly know $\left\{ \xH(t')\right \}_{t'=1}^{\tau}$.
\end{itemize}

We investigate this channel under the above set of two assumptions which is referred to as \emph{delayed CSIT} assumption. Let $\xR \Def (R_1,R_2,\cdots,R_K)\in (\mathbb{R^+})^K$ be a $K$-tuple of rates corresponding to the transmitter-receiver pairs (TX$_1$,RX$_1$), (TX$_2$,RX$_2$), $\cdots$, (TX$_K$,RX$_K$). A $(2^{\tau \xR},\tau)$ code of block length $\tau$ and rate $\xR$ with delayed CSIT consists of $K$ sets of encoding functions $\{\varphi^{[i]}_{t,\tau}\}_{1\leq t\leq \tau}$, $1\leq i \leq K$,
\begin{equation}
\begin{split}
\varphi_{t,\tau}^{[i]}&: \xxW^{[i]}\times \mathbb{C}^{K\times K \times (t-1)} \to \mathbb{C},\\
x_i(t)&=\varphi_{t,\tau}^{[i]} ( W^{[i]},\left\{\xH(t')\right \}_{t'=1}^{t-1} ),\quad 1\leq t \leq \tau,
\end{split}
\end{equation}
satisfying the power constraint of \cref{Eq:PowerConstraint}, together with $K$ decoding functions $\psi^{[i]}_\tau$, $1\leq i \leq K$,
\begin{equation}
\begin{split}
\psi^{[i]}_\tau& :\mathbb{C}^{\tau} \times \mathbb{C}^{K\times K \times \tau }\to \xxW^{[i]}, \\
\hat{W}^{[i]}_\tau&=\psi^{[i]}_\tau ( \left\{y_i(t),\xH(t)\right\}_{t=1}^\tau ).
\end{split}
\end{equation}
All encoding and decoding functions are revealed to all transmitters and receivers before the transmission begins. Defining the average probability of error at RX$_i$, $1\leq i \leq K$, for this code as
\begin{align}
P_{e,\tau}^{[i]}\Def \mathbb{P}\left(\hat{W}^{[i]}_\tau\neq W^{[i]}\right)=\frac{1}{2^{\tau R_i}}\sum_{w=1}^{2^{\tau R_i}}\mathbb{P}\left(\hat{W}^{[i]}_\tau\neq W^{[i]}| W^{[i]}=w\right), \nonumber
\end{align}
the average probability of error of this code is given by
\begin{equation}
P_{e,\tau} \Def \max_{1\leq i \leq K} P_{e,\tau}^{[i]}.
\end{equation}

For a given power constraint $P$, a rate tuple $\xR(P)$ is said to be achievable if there exists a sequence $\left\{(2^{\tau\xR(P)},\tau)\right\}_{\tau=1}^\infty$ of codes, for which $\lim_{\tau \to \infty}P_{e,\tau}=0$. The closure of the set of all achievable rate tuples $\xR(P)$ is called the capacity region of this channel with delayed CSIT and power constraint $P$ and is denoted by $\xxC(P)$. 

We consider this channel with $P \rightarrow \infty$, and define an \emph{achievable sum degrees of freedom} (simply called achievable degrees of freedom, or achievable DoF) as follows: $\DoFa_1^\text{IC}(K)$ is called an achievable DoF if and only if there exists an achievable rate tuple $\xR(P)$ such that
\begin{align}
\sum_{i=1}^K R_i(P)= \DoFa_1^\text{IC}(K) \times \log_2 P + o(\log_2 P).
\end{align}
The supremum of all achievable DoFs is called the DoF of this channel with delayed CSIT, and is denoted by $\DoF_1^\text{IC}(K)$.

We indeed consider a more general transmission setup as follows. Fix an integer $m$, $1\leq m \leq K$. Define $\xxS_m$ as a subset of cardinality $m$ of $\{1,2,\cdots,K\}$. Obviously, $\xxS_K= \{1,2,\cdots,K\}$. For any subset $\xxS_m \subset \xxS_K$ and any $i\in \xxS_m$, TX$_i$ wishes to communicate a common message $W^{[i|\xxS_m]} \in \xxW^{[i|\xxS_m]}$ of rate $R^{[i|\xxS_m]}$ to all receivers RX$_j$, $j\in \xxS_m$. We call $W^{[i|\xxS_m]}$ an order-$m$ message. The case $m=1$ represents the interference channel with private messages as described earlier. The codes, probabilities of error, achievable rates, capacity region, and degrees of freedom are similarly defined as before, now for a $K\binom{K}{m-1}$-tuple of rates. For any $1\leq m\leq K$, the DoF (resp.\ achievable DoF) of transmission of order-$m$ messages over the SISO IC with delayed CSIT is denoted by $\DoF_m^\text{IC}(K)$ (resp.\ $\DoFa_m^\text{IC}(K)$).

For any integers $M,K\geq2$, the discrete-time $M\times K$ SISO AWGN X channel is defined by a set of $M$ transmitters TX$_i$, ${1\leq i \leq M}$, and $K$ receivers RX$_j$, ${1\leq j\leq K}$,  with the input-output relationship
\begin{equation}
y_j (t)= \sum_{i=1}^M h_{ji}(t) x_i(t) + z_j(t), \hspace{5mm}1\leq j \leq K,
\end{equation}
where the parameters are defined as in \cref{Eq:IC-InputOutput}. The channel input is similarly subject to the average power constraint of \cref{Eq:PowerConstraint}. The delayed CSIT assumption is defined exactly as in the IC case. Fix an integer $m$, $1\leq m \leq K$. For any subset $\xxS_m \subset \xxS_K$ and any $i\in \xxS_M$, TX$_i$ wishes to communicate a common message $W^{[i|\xxS_m]} \in \xxW^{[i|\xxS_m]}$ of rate $R^{[i|\xxS_m]}$ to all receivers RX$_j$, $j\in \xxS_m$. The case $m=1$ corresponds to the X channel with private messages.

The achievable rates, capacity region, and degrees of freedom are similarly defined, now for an $M\binom{K}{m}$-tuple of rates. The DoF (resp.\ achievable DoF) of this channel under delayed CSIT assumption is denoted by $\DoF_m^\text{X}(M,K)$ (resp.\ $\DoFa_m^\text{X}(M,K)$) for $1\leq m \leq K$.

\section{Prior Art}
\label{Sec:PriorArt}
In order to achieve DoF gains in the $K$-user MISO broadcast channel with delayed CSIT, Maddah-ali et al. in \cite{maddah2012completely} proposed a multi-phase transmission scheme. The scheme starts with transmission of information symbols (a.k.a. order-$1$ symbols) in phase $1$ and continues phase by phase up to phase $K$. They defined an order-$m$ symbol as a piece of information available at the transmitter which is desired to be decoded by $m$ out of the $K$ receivers. The transmission in each phase is designed based on the following key ideas:
\begin{enumerate}[(i)]
\item \label{Item:BC_scheduling} \emph{Centralized Partial Interference Management via Receiver Scheduling}: During phase $m$, $1\leq m \leq K$, the transmitter transmits order-$m$ symbols on ``$m$ receivers per time slot'' basis. This implies that each received piece of information at each receiver is either purely desired (interference-free) or purely undesired (interference).
\item \label{Item:BC_RIA} \emph{Centralized Retrospective Interference Alignment (IA) via Retransmission}: At the end of phase $m$, $1\leq m \leq K-1$, since the transmitter has access to \emph{both} past CSI and transmitted order-$m$ symbols, it perfectly knows the whole past interference terms at each receiver. On the other hand, each interference term at a (non-scheduled) receiver is a useful piece of information for $m$ specific scheduled receivers about their desired order-$m$ symbols. Therefore, retransmission of each of such interference terms not only aligns the past interference at one receiver, but also provides another $m$ receivers with a desired piece of information.
\item \label{Item:BC_HigherOrder} \emph{Centralized Higher-Order Symbol Generation}: In conjunction with observation (\ref{Item:BC_RIA}), the set of all past interference terms can be partitioned into groups of $m+1$ interference terms (each available at one receiver) such that all quantities in a group are either known or desired by each of the $m+1$ receivers in that group. Therefore, delivering a linear combination of them to the corresponding $m+1$ receivers yields the retrospective IA according to observation (\ref{Item:BC_RIA}). Such linear combinations are considered as order-$(m+1)$ symbols to be transmitted during phase $m+1$.
\end{enumerate}

Using the above ideas, the transmission is done phase by phase up to phase $K$, wherein order-$K$ symbols are delivered to all receivers. Moreover, by providing a matching upper bound, the authors in \cite{maddah2012completely} proved DoF optimality of their multi-phase scheme.

Unlike the MISO broadcast channel, in SISO networks with distributed transmitters such as interference and X channels, there is a fundamental constraint: each transmitter has access only to its own information symbols. In the context of delayed CSIT, this constraint turns into a major bottleneck in using knowledge of the past CSI at transmitters. More specifically, 
\begin{itemize}
\item in both interference and X channels, in contrast to observation (\ref{Item:BC_RIA}) above,  a transmitter in spite of having access to the whole past CSI, cannot obtain the whole past interference at a receiver when the interference is due to more than one interferer. 
\item in the interference channel, in contrast to observation (\ref{Item:BC_scheduling}), the ``one receiver per time slot'' scheduling during phase $1$ immediately prevents the transmitters from achieving more than one DoF, since only one single-antenna transmitter has symbols to transmit for a scheduled receiver.
\end{itemize}
Nevertheless, the possibility of some level of retrospective interference alignment for the $3$-user IC and $2\times 2$ X channel with delayed CSIT was first demonstrated in \cite{maleki2012retrospective} by achieving more than $1$ DoF for both channels. In \cite{maleki2012retrospective}, it was proposed to partially alleviate the aforementioned bottlenecks based on an idea which we call \emph{Partial Interference Nulling (PIN)}\footnote{It is notable that in\cite{maleki2012retrospective}, the authors present a different analysis for their proposed transmission schemes. However, the partial interference nulling approach presented here provides a unified framework to compare the ideas proposed in \cite{maleki2012retrospective,Ghasemi2011Xchannel} with those proposed in this paper.}:

\begin{itemize}
\item \emph{PIN via Redundancy Transmission}: In both $3$-user IC and $2\times 2$ X channel, if all transmitters simultaneously transmit information symbols to all receivers, then each receiver in each time slot will experience interference from \emph{two} interferers. Therefore, if each transmitter transmits some redundancy together with the information symbols, each unintended receiver will be able to null out the interference from one of the interferers by performing a linear processing on its received linear combinations and obtain new linear combinations, each of which contains interference only due to one interferer. Now, by having access to the past CSI, each of these remaining pieces of interference can be reconstructed by its corresponding interferer.
\end{itemize}
\begin{remark}
The ``redundancy'' in our interpretation is redundancy per scheduled information flow. For instance, in the $2\times 2$ X channel, if each transmitter transmits $3$ linear combinations of $4$ information symbols for both receivers, \ie $2$ symbols per receiver, during $3$ time slots, it is a redundancy transmission. Indeed, for each transmitter-receiver pair, $3$ linear combinations of $2$ symbols have been transmitted.
\end{remark}

Using the above idea, they proposed two-phase schemes which employ PIN during phase $1$. Then, during phase $2$, the retrospective IA is achieved in a distributed manner as follows.

\begin{itemize}
\item \emph{Distributed Retrospective IA in the $2\times 2$ X Channel}: Each transmitter retransmits its past interference term remaining at each receiver, and hence, aligns the interference at the unintended receiver while providing a desired piece of information to the intended receiver.

\item \emph{Distributed Retrospective IA in the $3$-user IC}: Each transmitter transmits a specific linear combination of its past interference terms remaining at its both unintended receivers. The transmitted linear combination is such that it aligns the remaining past interference at \emph{both} unintended receivers while providing a desired piece of information to the intended receiver.
\end{itemize}

Subsequently, Ghasemi et al. in \cite{Ghasemi2011Xchannel} proposed a two-phase transmission scheme for the $K\times K$ X channel which in the case of $K=2$ achieves a higher DoF than that of \cite{maleki2012retrospective}. Their scheme is based on the following ideas.
\begin{itemize}
\item \emph{Distributed Partial Interference Management via User Scheduling in Phase $1$}: Since each transmitter has an information symbol for each receiver, the users are scheduled on ``all transmitters/one receiver per time slot'' basis during phase $1$. In this way, one receiver (the scheduled receiver) in each time slot receives interference-free information. For $K=2$, this is in contrast to ``all transmitters/all receivers per time slot'' scheduling in \cite{maleki2012retrospective}.
\item \emph{PIN in Phase $1$}: A redundancy transmission is designed in conjunction with the mentioned user scheduling. As a result, each receiver during its non-scheduled time slots performs PIN and obtains a piece of interference, which is desired by the scheduled receiver and also can be regenerated by one of the transmitters using delayed CSIT. 
\item \emph{Distributed Retrospective IA in Phase $2$}: Each of the mentioned past interference terms is retransmitted for its intended receiver by its corresponding transmitter, and thus, the retrospective IA is attained.
\end{itemize}

\section{Main Contributions and Results}
\label{Sec:MainResultsDiscussions}

\subsection{Main Contributions}
\label{Sec:MainContributions}

In the following, we highlight the key ideas which constitute the main contributions of this paper and are the main building blocks of our $K$-phase transmission schemes for both the $K$-user IC and $2\times K$ X channel.

\begin{enumerate}[(a)]
\item \emph{Multi-phase PIN and Retrospective IA}: In the context of delayed CSIT, any PIN approach, which is realized using redundancy transmission, involves an inherent trade-off: On the one hand, transmission of any amount of redundancy is equivalent to DoF loss. On the other hand, an appropriate redundancy can help the receivers to partially null the received interference and achieve a more effective retrospective IA, which results in DoF gain.

Recognizing the mentioned trade-off, we propose to gradually transmit the redundancy over several phases, namely $K-1$ phases. The amount of transmitted redundancy in each phase is designed such that each receiver is enabled to null out the received interference from one interferer, if any. The remaining piece(s) of interference are then retransmitted one by one by their corresponding interferers to achieve retrospective IA. Let us compare this idea with the prior art.
\begin{itemize}
\item \emph{Comparison with \cite{maddah2012completely}}: No PIN is done in \cite{maddah2012completely}.
\item \emph{Comparison with \cite{maleki2012retrospective}}: For the case of $3$-user IC, our approach is in contrast to the single-phase PIN approach of \cite{maleki2012retrospective}, wherein the entire PIN is accomplished during phase $1$. Indeed, they achieve the retrospective IA over a single phase (phase $2$) at the expense of more redundancy transmitted in phase $1$. In contrast, we propose a $3$-phase scheme which distributes the amount of redundancy over phases $1$ and $2$, and achieves the retrospective IA over phases $2$ and $3$. This approach yields a better trade-off and achieves a greater DoF than that of \cite{maleki2012retrospective}.
\item \emph{Comparison with \cite{Ghasemi2011Xchannel}}: For the case of $2\times K$ X channel with $K>2$, our approach is in contrast to single-phase PIN approach of \cite{Ghasemi2011Xchannel}, wherein the entire PIN is accomplished during phase $1$.
\end{itemize}

\item \emph{Distributed Partial Interference Management via User Scheduling}: Except for phase $1$ of our scheme for the $K$-user interference channel, in all phases up to phase $K-1$ of our schemes for both channels, we perform distributed partial interference management by scheduling specific transmitters/receivers per time slot as follows. For both channels, during phase $m$ ($m\geq 2$ for the IC), the transmitters and receivers are respectively scheduled on ``one pair of transmitters per time slot''  and ``one $m$-tuple of receivers per time slot'' bases. This scheduling ensures interference-free reception of information at the $m$ scheduled receivers. The following compares our approach with the prior art.
\begin{itemize}
\item \emph{Comparison with \cite{maddah2012completely}}: The main novelty of our interference management approach is due to its transmitter scheduling. In fact, in the broadcast channel, studied in \cite{maddah2012completely}, the single transmitter is always trivially scheduled.
\item \emph{Comparison with \cite{maleki2012retrospective}}: Their schemes have no interference management via user scheduling.
\item \emph{Comparison with \cite{Ghasemi2011Xchannel}}: During phase $1$ of our scheme for the $X$ channel, we perform a ``two transmitters per time slot'' scheduling whereas \cite{Ghasemi2011Xchannel} performs a ``$K$ transmitters per time slot'' scheduling.
\end{itemize}

\item \emph{Distributed Higher-Order Symbol Generation}: We propose a distributed higher-order symbol generation which has the following novelties compared with \cite{maddah2012completely}.
\begin{itemize}
\item In both the $K$-user IC and $2\times K$ X channel, we generate the higher order symbols in a distributed fashion rather than centralized approach of \cite{maddah2012completely}. More specifically, in partitioning the set of all remaining past interference terms for generation of higher order symbols, we ensure that each generated higher order symbol is reconstructable by one of the transmitters (cf.\ observation (\ref{Item:BC_HigherOrder}) of \cref{Sec:PriorArt}).
\item In the $K$-user IC, the generation of order-$2$ symbols at the end of phase $1$ of our scheme is totally different than that of \cite{maddah2012completely}, since our receiver scheduling approach is different.
\item In the $K$-user IC, our scheme involves generation and transmission of a new type of symbols, called order-$(1,m)$ symbols, which are not defined or dealt with in \cite{maddah2012completely}.
\end{itemize}
\end{enumerate}

\subsection{Main Results}
\label{Sec:MainResults}
The main results of this paper are formulated in the following two theorems, whose proofs are provided in \cref{Sec:SISO-IC,Sec:SISO-X}.
\begin{theorem}
\label{Th:DoF-IC}
In the $K$-user SISO interference channel with delayed CSIT and $K\geq2$, $\DoFa_1^\text{IC}(K)$ degrees of freedom is achievable almost surely, where $\DoFa_1^\text{IC}(K)$ is obtained by
\begin{align}
\DoFa_1^\text{IC}(K)=\left[1-\frac{K-2}{K(K-1)^2}-\frac{K-2}{K-1}A_2(K)\right]^{-1}, \label{Eq:DoF(K)-IC}
\end{align}
and $A_2(K)$ is given by
\begin{align}
A_2(K)&\Def-\frac{(K-2)(K-3)}{4\left[4(K-2)^2-1\right]}+\sum_{\ell_1=0}^{K-3}\frac{(K-\ell_1-1)(3\ell_1^2+\ell_1-1)}{2(K-\ell_1)(4\ell_1^2-1)}\prod_{\ell_2=\ell_1+1}^{K-2}\frac{\ell_2}{2\ell_2+1}.  \label{Eq:A_2(K)}
\end{align}

Moreover, for $2\leq m \leq K$, $\DoFa_{m}^\text{IC}(K)$ degrees of freedom is achievable in transmission of order-$m$ messages, where
\begin{align}
\DoFa_{m}^\text{IC}(K)=\Bigg[1+\frac{(K-m)(K-m-1)}{2m[4(K-m)^2-1]}-\sum_{\ell_1=0}^{K-m-1}\frac{(K-m-\ell_1{+}1)(3\ell_1^2{+}\ell_1-1)}{2(K-\ell_1)(4\ell_1^2-1)}\prod_{j=\ell_1+1}^{K-m}\frac{\ell_2}{2\ell_2+1}\Bigg]^{-1}. \label{Eq:DoF_m(K)-IC}
\end{align}

\end{theorem}

\begin{theorem}
\label{Th:DoF-X}
In the $2\times K$ SISO X channel with delayed CSIT and $K\geq2$, $\DoFa_1^\text{X}(2,K)$ degrees of freedom is achievable almost surely, where 
\begin{align}
\DoFa_1^\text{X}(2,K)=\left[ 1-\sum_{\ell_1=0}^{K-2}\frac{(K-1-\ell_1)(\ell_1+1)}{(K-\ell_1)(2\ell_1+1)}\prod_{\ell_2=\ell_1+1}^{K-1}\frac{\ell_2}{2\ell_2+1}\right]^{-1}. \label{Eq:DoF(K)-X}
\end{align}

More generally, for $2\leq m \leq K$, $\DoFa_{m}^\text{X}(2,K)$ degrees of freedom is achievable in transmission of order-$m$ messages, where
\begin{align}
\DoFa_{m}^\text{X}(2,K)= \left[ 1-\sum_{\ell_1=0}^{K-m-1}\frac{(K-m-\ell_1)(\ell_1+1)}{(K-\ell_1)(2\ell_1+1)}\prod_{\ell_2=\ell_1+1}^{K-m}\frac{\ell_2}{2\ell_2+1}\right]^{-1}. \label{Eq:DoF_m(K)-X} 
\end{align}

\end{theorem}

\begin{remark}
Using scaled versions of the schemes proposed in \cref{Sec:SISO-IC,Sec:SISO-X}, $N\DoFa_1^\text{IC}(K)$ and  $N\DoFa_1^\text{X}(2,K)$ are achievable in the $K$-user MIMO IC and $2\times K$ MIMO X channel, respectively, with $N$ antennas available at each node and with delayed CSIT, where $\DoFa_1^\text{IC}(K)$ and $\DoFa_1^\text{X}(2,K)$ are given by \cref{Eq:DoF(K)-IC,Eq:DoF(K)-X}, respectively.
\end{remark}
\subsection{Numerical Comparison}
\label{Sec:NumericalComparison}

\begin{figure}
\centering
\includegraphics[width=12cm]{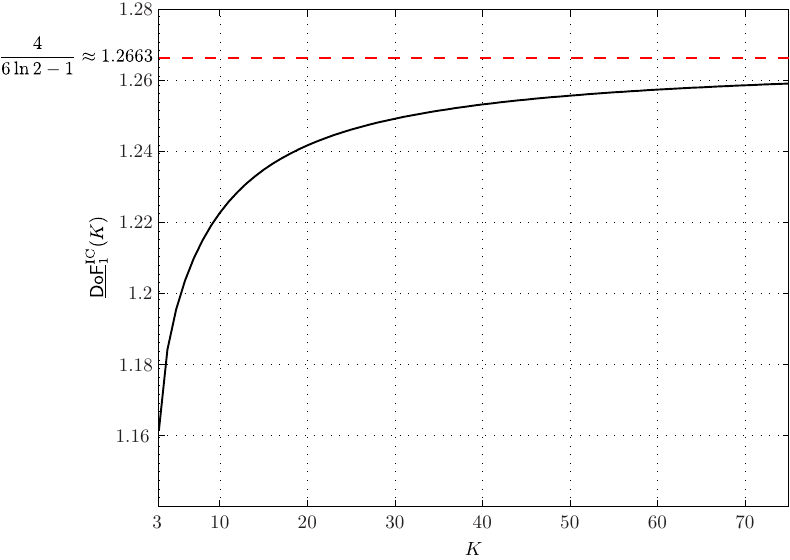}
\caption{Our achievable DoF for the $K$-user SISO interference channel with delayed CSIT and $3\leq K \leq 75$.}
\label{Fig:DoF(K)-IC}
\end{figure}

\begin{figure}
\centering
\includegraphics[width=12cm]{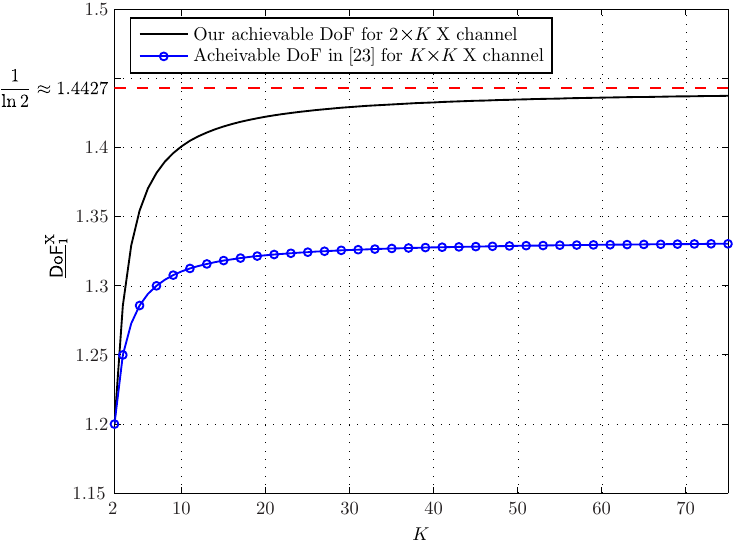}
\caption{Our achievable DoF for the SISO X channel with delayed CSIT and $2\leq K \leq 75$.}
\label{Fig:DoF(K)-X}
\end{figure}

Our achievable DoFs for the $K$-user SISO IC and $2\times K$ SISO X channel with private messages and delayed CSIT are plotted in \cref{Fig:DoF(K)-IC,Fig:DoF(K)-X} for $2\leq K \leq 75$, respectively. For the sake of comparison, the achievable DoF reported in \cite{Ghasemi2011Xchannel} for the $K\times K$ SISO X channel with delayed CSIT is also plotted in \cref{Fig:DoF(K)-X}. For $K\geq 3$, our achievable DoF for the $2\times K$ X channel, \ie $\DoFa_1^\text{X}(2,K)$ presented in \cref{Th:DoF-X}, is strictly greater than $\frac{4}{3}-\frac{2}{3(3K-1)}$ DoF achieved in \cite{Ghasemi2011Xchannel} for the $K \times K$ X channel. It is proved in Appendix \ref{App:DoF-Limit} that, as $K\to \infty$, the achievable DoFs approach limiting values of $\frac{4}{6\ln 2-1}\approx 1.2663$ and $\frac{1}{\ln2}\approx 1.4427$ for the IC and X channel, respectively. \Cref{Tbl:DoF_IC,Tbl:DoF_X} list our achievable DoFs for the $K$-user IC and $2\times K$ X channel with delayed CSIT and $2\leq K \leq 5$.

\subsection{Conjecture on DoF Scaling}
\label{Sec:Conjecture}

In the lack of tight upper bounds, no optimality argument can be made about our achievable DoFs. However, we conjecture that, under the delayed CSIT assumption, the DoF of both the $K$-user IC and $M\times K$ X channel in i.i.d.\ fading environment, for any $M,K\geq2$, is bounded above by a constant, \ie it does not scale with number of users. In what follows, we provide some insights on this conjecture.

Transmission in any channel with delayed CSIT possesses a ``transmit-retransmit'' nature: Since the transmitters do not have the current CSI, they inevitably transmit some information regardless of the current CSI. Then, using the knowledge of past CSI, they access and retransmit (part of) the past interference. Now, we have the following observations in channels with distributed transmitters.
\begin{itemize}
\item Since each transmitter is equipped with a single antenna, the number of independent pieces of information transmitted over the channel per time slot is upper bounded by the number of active transmitters per time slot. 
\item During a ``transmit'' phase, the transmitters cannot exploit the knowledge of current CSI to do any instantaneous interference alignment. Hence, the number of interference parts in the signal received by each receiver is proportional to the number of active transmitters per time slot. Indeed, we conjecture that the former always falls within a constant gap below the latter. This considers using PIN or any no-CSIT based interference mitigation technique.
\item Since the transmitters are distributed, the number of past interference parts which need to be retransmitted is within a constant gap of the number of the interference parts in each received signal at the receiver side.
\end{itemize}

Summarizing the above observations, the number of interference quantities per time slot of a ``transmit'' phase which need to be subsequently retransmitted is conjectured to be within a constant gap below the number of transmitted independent symbols per time slot of that phase. Denoting the latter number with $D(K)$ and the constant gap with $\lambda$, we have the upper bound
\begin{align}
\DoF(K) \leq \frac{D(K)}{1+\frac{D(K)-\lambda}{\DoF(K)}}, \label{Eq:DoF_Conjecture_Bound}
\end{align}
for the DoF, where we have used the fact that DoF of delivering the past interference parts to their corresponding subsets of receivers does not exceed the channel DoF. Then, \cref{Eq:DoF_Conjecture_Bound} immediately yields $\DoF(K)\leq \lambda$.

Finally, it is worth mentioning that the only known upper bounds in the literature on the DoF of channels with delayed CSIT are due to \cite{maddah2012completely} for the $K$-user MISO broadcast channel, and \cite{Vaze2012IC_DCSIT} for the two-user MIMO interference channel. While those upper bounds proved to be tight for their corresponding channels, their extension to the channels considered in this paper does not seem to be possible. Indeed, the idea in \cite{maddah2012completely} is to enhance the MISO broadcast channel to a MIMO physically degraded broadcast channel and use the fact that feedback does not increase the capacity of a physically degraded broadcast channel according to \cite{gamal1978feedback}. This result does not have a counterpart in the interference or X channels. Also, the fundamental difference between the two-user MIMO interference channel of \cite{Vaze2012IC_DCSIT} and the $K$-user channels considered in this paper is that in the two-user interference channel, the interference at each receiver is due to only one interferer. This is in contrast to multi-interferer nature of the $K$-user IC and X channel.

\begin{table*}
\caption{Achievable DoFs for the SISO interference channel with delayed CSIT}
\begin{center}
\begin{tabular}{ccccc}
\toprule
$K$ & $2$ & $3$ & $4$ & $5$ \\
\midrule
\midrule
Our achievable DoF for the $K$-user IC & $1$ & $\frac{36}{31}$ & $\frac{45}{38}$ & $\frac{1400}{1171}$ \\
\bottomrule
\end{tabular}
\end{center}
\label{Tbl:DoF_IC}
\end{table*}

\begin{table*}
\caption{Achievable DoFs for the SISO X channel with delayed CSIT}
\begin{center}
\begin{tabular}{ccccc}
\toprule
$K$ & $2$ & $3$ & $4$ & $5$\\
\midrule
\midrule
Our achievable DoF for the $2\times K$ X channel & $\frac{6}{5}$ & $\frac{9}{7}$ & $\frac{105}{79}$ & $\frac{1575}{1163}$\\
\midrule
Achievable DoF in \cite{Ghasemi2011Xchannel} for the $K\times K$ X channel & $\frac{6}{5}$ & $\frac{5}{4}$ & $\frac{14}{11}$ & $\frac{9}{7}$\\
\bottomrule
\end{tabular}
\end{center}
\label{Tbl:DoF_X}
\end{table*}

\section{Proof of Theorem 1}
\label{Sec:SISO-IC}
In this section, we prove that $\DoFa_m^\text{IC}(K)$, $1\leq m \leq K$, stated in \cref{Th:DoF-IC} can be achieved in the $K$-user SISO IC with delayed CSIT. To this end, we first elaborate on our achievable scheme for the case of $K=3$. We then propose our transmission scheme for the general $K$-user setting. 

Before proceeding with the transmission schemes, let us define some notations which will be widely used throughout the paper. We use $u^{[i|\xxS_m;\xxS_n]}$ to denote a symbol which is
\begin{itemize}
\item available at TX$_i$,
\item available at RX$_j$, for every $j\in \xxS_n$,
\item intended to be decoded at RX$_k$, for every $k\in \xxS_m$.
\end{itemize}
We refer to $u^{[i|\xxS_m;\xxS_n]}$ as an $(\xxS_m;\xxS_n)$-symbol available at TX$_i$. The \emph{order} of symbol $u^{[i|\xxS_m;\xxS_n]}$ is defined as the ordered pair $(m,n)$ containing the cardinalities of $\xxS_m$ and $\xxS_n$, respectively. For instance, $u^{[2|1,5;3]}$ is a $(1,5;3)$-symbol of order $(2,1)$ which is available at TX$_2$ and RX$_3$, and is intended to be decoded at both RX$_1$ and RX$_5$, where the set braces ``$\{$'' and ``$\}$'' have been omitted to avoid cumbersome notations. For ease of notation, a symbol $u^{[i|\xxS_m;\xxS_n]}$ with $\xxS_n=\{\}$ is denoted by $u^{[i|\xxS_m]}$ and is called an $\xxS_m$-symbol of order $m$. 

\subsection{The $3$-user SISO Interference Channel}
\label{Sec:SISO-IC3}
\begin{figure}
\centering
\includegraphics[scale=1]{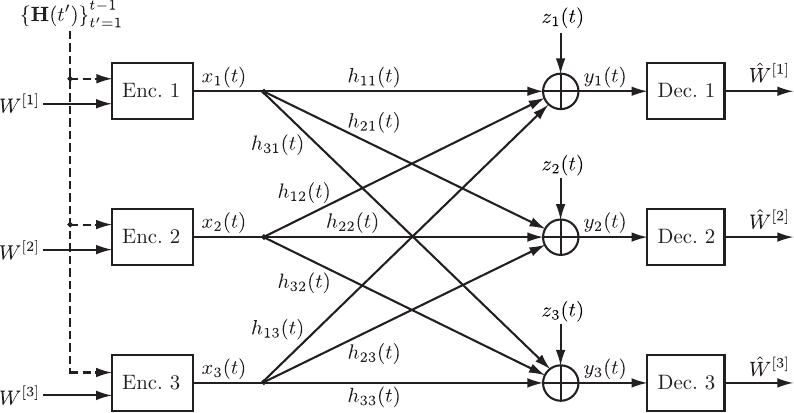}
\caption{The $3$-user SISO interference channel with delayed CSIT.}
\label{Fig:3-User-IC}
\end{figure}

Consider the $3$-user SISO IC with delayed CSI available at the transmitters as depicted in \cref{Fig:3-User-IC}.
In order to achieve ${\DoFa_1^\text{IC}(3)=\frac{36}{31}}$, suggested by \cref{Eq:DoF(K)-IC}, transmission is accomplished in three distinct phases. The fresh information symbols are fed to the channel in the first phase. During the remaining phases, extra linear combinations are delivered to the receivers in such a way that the interference is properly aligned at each receiver. At the end of transmission scheme, the receivers are left with the desired number of equations in terms of their respective information symbols.

It is important to point out that we will use several random coefficients during the transmission scheme to construct and transmit different channel input symbols. These coefficients are randomly generated and revealed to all transmitters and receivers before the beginning of communication. The transmission phases are described in detail as follows.

\vspace{3mm}
$\bullet$ \underline{\textit{Phase $1$ ($3$-user IC)}:}
\vspace{3mm}

This phase takes $5$ time slots, during which each transmitter feeds $4$ fresh information symbols to the channel. Since in the interference channel there exists no $u^{[i|j]}$ for ${j\neq i}$, we simply use $u^{[i]}$ instead of $u^{[i|i]}$. 

\emph{Redundancy Transmission}: Let ${\xu^{[i]} \Def [u^{[i]}_{1},u^{[i]}_{2},u^{[i]}_{3},u^{[i]}_{4}]^T}$ denote the vector containing the information symbols of TX$_i$, ${1\leq i \leq 3}$. In each time slot, each transmitter transmits a random linear combination of its $4$ information symbols. Let 
\begin{align}
\xc^{[i]}(t) \Def \left[c^{[i]}_{1}(t),c^{[i]}_{2}(t),c^{[i]}_{3}(t),c^{[i]}_{4}(t)\right]^T, \hspace{5mm}1\leq t \leq 5, \nonumber
\end{align}
denote the vector containing the random coefficients of the linear combination transmitted by TX$_i$, ${1\leq i \leq 3}$, over time slot $t$, \ie ${x_i(t)=\left(\xc^{[i]}(t)\right)^T\xu^{[i]}}$. Ignoring the noise terms at receivers, the received signal at RX$_j$ in time slot $t$ is equal to
\begin{align}
y_j(t)&=h_{j1}(t)x_1(t)+h_{j2}(t)x_2(t)+h_{j3}(t)x_3(t) \nonumber \\
& = h_{j1}(t)\left(\xc^{[1]}(t)\right)^{T} \xu^{[1]}+h_{j2}(t)\left(\xc^{[2]}(t)\right)^T \xu^{[2]}+h_{j3}(t)\left(\xc^{[3]}(t)\right)^T \xu^{[3]},\hspace{9mm}1\leq j \leq 3.
\end{align}
Therefore, by the end of phase $1$, RX$_j$ obtains the system of linear equations
\begin{align}
\xy_j = \xD_{j1}\xC^{[1]}\xu^{[1]}+\xD_{j2}\xC^{[2]}\xu^{[2]}+\xD_{j3}\xC^{[3]}\xu^{[3]}, \label{Eq:y-vector}
\end{align}
in terms of \emph{all} transmitted information symbols, where $\xy_j$ is the vector of received symbols at RX$_j$ during $5$ time slots, $\xD_{ji}$ is the $5\times5$ diagonal matrix containing $h_{ji}(t)$, $1\leq t \leq 5$, on its main diagonal, and $\xC^{[i]}$ is a $5\times4$ matrix containing the random coefficients employed by TX$_i$ during these $5$ time slots,
\begin{align}
\xC^{[i]} \Def \left[\xc^{[i]}(1)|\xc^{[i]}(2)|\cdots|\xc^{[i]}(5)\right]^T, \hspace{5mm}1\leq i \leq 3.
\end{align}

\emph{PIN}: Since the elements of $\xC^{[i]}$ and the elements of the diagonal of $\xD_{ji}$ are i.i.d., both matrices are full rank almost surely. Thereby, since $\xC^{[i]}$ and $\xD_{ji}$ are independent of each other, their product is also full rank, \ie ${\rank(\xQ_{ji})=4}$, where $\xQ_{ji} \Def \xD_{ji}\xC^{[i]}$, $1\leq i,j \leq 3$. Since $\xQ_{ji}$ is a full rank $5\times4$ matrix, its \emph{left null space} is one dimensional almost surely. As a result, for each $(i,j)$, ${1\leq i,j \leq 3}$, there exists a nonzero vector ${\xomega_{ji}=[\omega_{ji1},\omega_{ji2},\omega_{ji3},\omega_{ji4},\omega_{ji5}]^T}$ such that
\begin{equation}
\label{Eq:omega}
\xQ^T_{ji} \xomega_{ji} = \mathbf{0}_{4\times1}, \qquad 1\leq i,j \leq 3.
\end{equation}

Note that by the end of phase $1$, all transmitters and receivers have access to $\xQ_{ji}$, and thus, to $\xomega_{ji}$, ${1\leq i,j \leq 3}$. Hence, using \cref{Eq:y-vector,Eq:omega}, RX$_1$ can null out $\xu^{[3]}$ from its received signal and obtain
\begin{align}
\xy_1^T \xomega_{13}&=(\xu^{[1]})^T \xQ_{11}^T \xomega_{13}+(\xu^{[2]})^T \xQ_{12}^T \xomega_{13}+(\xu^{[3]})^T \overbrace{\xQ_{13}^T \xomega_{13}}^\mathbf{0} \nonumber \\
&= (\xu^{[1]})^T \xQ_{11}^T \xomega_{13}+(\xu^{[2]})^T \xQ_{12}^T \xomega_{13}, \label{Eq:RX1-1}
\end{align}
and null out $\xu^{[2]}$ to obtain
\begin{align}
\xy_1^T \xomega_{12}&=(\xu^{[1]})^T \xQ_{11}^T \xomega_{12}+(\xu^{[2]})^T \overbrace{\xQ_{12}^T \xomega_{12}}^\mathbf{0}+(\xu^{[3]})^T \xQ_{13}^T \xomega_{12}\nonumber \\
&=(\xu^{[1]})^T \xQ_{11}^T \xomega_{12}+(\xu^{[3]})^T \xQ_{13}^T \xomega_{12}. \label{Eq:RX1-2}
\end{align}
Similarly, RX$_2$ can obtain
\begin{align}
\xy_2^T \xomega_{21}&=(\xu^{[2]})^T \xQ_{22}^T \xomega_{21}+(\xu^{[3]})^T \xQ_{23}^T \xomega_{21}, \label{Eq:RX2-1}\\
\xy_2^T \xomega_{23}&=(\xu^{[2]})^T \xQ_{22}^T \xomega_{23}+(\xu^{[1]})^T \xQ_{21}^T \xomega_{23}, \label{Eq:RX2-2}
\end{align}
and RX$_3$ can obtain
\begin{align}
\xy_3^T \xomega_{31}&=(\xu^{[3]})^T \xQ_{33}^T \xomega_{31}+(\xu^{[2]})^T \xQ_{32}^T \xomega_{31}, \label{Eq:RX3-1}\\
\xy_3^T \xomega_{32}&=(\xu^{[3]})^T \xQ_{33}^T \xomega_{32}+(\xu^{[1]})^T \xQ_{31}^T \xomega_{32}. \label{Eq:RX3-2}
\end{align}

\emph{Order-$2$ Symbol Generation}: If we deliver $(\xu^{[1]})^T \xQ_{21}^T \xomega_{23}$, $(\xu^{[2]})^T \xQ_{12}^T \xomega_{13}$, $(\xu^{[1]})^T \xQ_{31}^T \xomega_{32}$, and $(\xu^{[3]})^T \xQ_{13}^T \xomega_{12}$ to RX$_1$, it can obtain enough equations to resolve its four desired information symbols as follows.
\begin{itemize}
\item $(\xu^{[1]})^T \xQ_{21}^T \xomega_{23}$ and $(\xu^{[1]})^T \xQ_{31}^T \xomega_{32}$ are two desired equations in terms of the $4\times1$ information vector $\xu^{[1]}$.
\item $(\xu^{[2]})^T \xQ_{12}^T \xomega_{13}$ can be subtracted from $\xy_1^T \xomega_{13}$ to yield $(\xu^{[1]})^T \xQ_{11}^T \xomega_{13}$, a desired equation in terms of $\xu^{[1]}$.
\item $(\xu^{[3]})^T \xQ_{13}^T \xomega_{12}$ can be subtracted from $\xy_1^T \xomega_{12}$ to yield $(\xu^{[1]})^T \xQ_{11}^T \xomega_{12}$, a desired equation in terms of $\xu^{[1]}$.
\end{itemize}
Therefore, RX$_1$ will have a system of four linear equations in terms of $4\times1$ information vector $\xu^{[1]}$, namely, $(\xu^{[1]})^T \xQ_{21}^T \xomega_{23}$, $(\xu^{[1]})^T \xQ_{31}^T \xomega_{32}$, $(\xu^{[1]})^T \xQ_{11}^T \xomega_{13}$, and $(\xu^{[1]})^T \xQ_{11}^T \xomega_{12}$. It is shown in Appendix \ref{App:LinearInependence} that these equations are linearly independent almost surely, and thus, RX$_1$ can solve them to obtain $\xu^{[1]}$. By a similar argument, having $(\xu^{[1]})^T \xQ_{21}^T \xomega_{23}$, $(\xu^{[2]})^T \xQ_{12}^T \xomega_{13}$, $(\xu^{[2]})^T \xQ_{32}^T \xomega_{31}$, and $(\xu^{[3]})^T \xQ_{23}^T \xomega_{21}$, RX$_2$ can obtain four linearly independent equations in terms of $\xu^{[2]}$, and so, it can solve them for $\xu^{[2]}$. Also, after providing RX$_3$ with $(\xu^{[1]})^T \xQ_{31}^T \xomega_{32}$, $(\xu^{[3]})^T \xQ_{13}^T \xomega_{12}$, $(\xu^{[2]})^T \xQ_{32}^T \xomega_{31}$, and $(\xu^{[3]})^T \xQ_{23}^T \xomega_{21}$, it can obtain enough equations to solve for $\xu^{[3]}$.

In summary, our goal in phase $2$ boils down to delivering $(\xu^{[1]})^T \xQ_{21}^T \xomega_{23}$ and $(\xu^{[2]})^T \xQ_{12}^T \xomega_{13}$ to both RX$_1$ and RX$_2$, delivering $(\xu^{[1]})^T \xQ_{31}^T \xomega_{32}$ and $(\xu^{[3]})^T \xQ_{13}^T \xomega_{12}$ to both RX$_1$ and RX$_3$, and delivering $(\xu^{[2]})^T \xQ_{32}^T \xomega_{31}$ and $(\xu^{[3]})^T \xQ_{23}^T \xomega_{21}$ to both RX$_2$ and RX$_3$. Therefore, the following order-$2$ symbols can be defined.
\begin{align}
u^{[1|1,2]}&\Def (\xu^{[1]})^T \xQ_{21}^T \xomega_{23}, \quad u^{[1|1,3]}\Def (\xu^{[1]})^T \xQ_{31}^T \xomega_{32}, \label{Eq:u^[1|1j]}\\
u^{[2|1,2]}&\Def (\xu^{[2]})^T \xQ_{12}^T \xomega_{13}, \quad u^{[2|2,3]}\Def (\xu^{[2]})^T \xQ_{32}^T \xomega_{31},\label{Eq:u^[2|2j]} \\
u^{[3|1,3]}&\Def (\xu^{[3]})^T \xQ_{13}^T \xomega_{12}, \quad u^{[3|2,3]}\Def (\xu^{[3]})^T \xQ_{23}^T \xomega_{21}. \label{Eq:u^[3|3j]}
\end{align}

\vspace{3mm}
$\bullet$  \underline{\textit{Phase $2$ ($3$-user IC)}:}
\vspace{3mm}

This phase takes $12$ time slots to transmit $18$ order-$2$ symbols generated in phase $1$. Since we generated only $6$ order-$2$ symbols in phase $1$, we simply repeat phase $1$ three times to obtain $18$ order-$2$ symbols required in phase $2$. This takes $3\times5=15$ time slots, and hence, phase $2$ begins at time slot $t=16$. Consequently, at the beginning of phase $2$, for any $\{i,j\}\subset\{1,2,3\}$, there are three order-$2$ symbols $u^{[i|i,j]}_{1}$ , $u^{[i|i,j]}_{2}$, and $u^{[i|i,j]}_{3}$ at TX$_i$ and  three order-$2$ symbols $u^{[j|i,j]}_{1}$ , $u^{[j|i,j]}_{2}$, and $u^{[j|i,j]}_{3}$ at TX$_j$. The transmission in phase $2$ is then carried out as follows.

\emph{Redundancy Transmission}: In the first time slot of phase $2$, TX$_1$ transmits a random linear combination of $u^{[1|1,2]}_{1}$ and $u^{[1|1,2]}_{2}$ while TX$_2$  transmits $u^{[2|1,2]}_{1}$. In the second time slot, TX$_1$ transmits another random linear combination of $u^{[1|1,2]}_{1}$ and $u^{[1|1,2]}_{2}$ while TX$_2$ repeats $u^{[2|1,2]}_{1}$. TX$_3$ is silent during these two time slots. Hence, each receiver obtains two linearly independent equations in terms of three $(1,2)$-symbols $u^{[1|1,2]}_{1}$, $u^{[1|1,2]}_{2}$, and $u^{[2|1,2]}_{1}$. As such, each of RX$_1$ and RX$_2$ requires an extra equation to resolve these three order-$2$ symbols. 

\emph{PIN}: Consider the linear combinations received at RX$_3$ during these two time slots, \ie $t=16,17$,
\begin{align}
y_{3}(t)&=h_{31}(t)x_{1}(t)+h_{32}(t)x_{2}(t)\nonumber \\
&=h_{31}(t)\left(\xc^{[1|1,2]}(t)\right)^{T}\xu^{[1|1,2]}+h_{32}(t)u^{[2|1,2]}_{1},
\end{align}
where $\xu^{[1|1,2]} \Def[u_{1}^{[1|1,2]},u_{2}^{[1|1,2]}]^{T}$, and 
\begin{align}
\xc^{[1|1,2]}(t)\Def\left[c_{1}^{[1|1,2]}(t),c_{2}^{[1|1,2]}(t)\right]^{T}
\end{align}
 is the vector of random coefficients employed by TX$_1$ in time slot $t$. Now, RX$_3$ can null out $u^{[2|1,2]}_{1}$ and form
\begin{align}
\frac{1}{h_{32}(16)}&y_{3}(16)-\frac{1}{h_{32}(17)}y_{3}(17)=\left[\frac{h_{31}(16)}{h_{32}(16)}\left(\xc^{[1|1,2]}(16)\right)^{T}-\frac{h_{31}(17)}{h_{32}(17)}\left(\xc^{[1|1,2]}(17)\right)^{T}\right]\xu^{[1|1,2]}, \nonumber
\end{align}
which is an equation solely in terms of the elements of $\xu^{[1|1,2]}$. This is the side information that RX$_3$ has about the order-$2$ symbols of RX$_1$ and RX$_2$, and can provide the extra equation required by both RX$_1$ and RX$_2$ to resolve their order-$2$ symbols. Based on our terminology, this quantity is denoted by $u^{[1|1,2;3]}$. 

The next two time slots are dedicated to the transmission of another three order-$2$ $(1,2)$-symbols. Now, the roles of TX$_1$ and TX$_2$ are exchanged. Specifically, during time slots ${t=18,19}$, TX$_2$ transmits two random linear combinations of $u^{[2|1,2]}_{2}$ and $u^{[2|1,2]}_{3}$ while TX$_1$ repeats the same symbol $u^{[1|1,2]}_{3}$. The side information $u^{[2|1,2;3]}$ is similarly formed at RX$_3$ by the end of these two time slots. 

Up to this point, we have transmitted $6$ order-$2$ $(1,2)$-symbols in $4$ time slots, and generated two pieces of side information at RX$_3$. Analogously, for each of receiver pairs $\{1,3\}$ and $\{2,3\}$, the above procedure can be repeated using their respective transmitters. Therefore, by spending another $2\times4=8$ time slots, we will transmit $2\times6=12$ order-$2$ symbols and generate the side information $u^{[2|2,3;1]}$ and $u^{[3|2,3;1]}$ at RX$_1$, and $u^{[1|1,3;2]}$ and $u^{[3|1,3;2]}$ at RX$_2$. Therefore, our goal is reduced to
\begin{itemize}
\item[(a)] delivering $u^{[1|1,2;3]}$ and $u^{[2|1,2;3]}$ to both RX$_1$ and RX$_2$,
\item[(b)] delivering $u^{[1|1,3;2]}$ and $u^{[3|1,3;2]}$ to both RX$_1$ and RX$_3$,
\item[(c)] delivering $u^{[2|2,3;1]}$ and $u^{[3|2,3;1]}$ to both RX$_2$ and RX$_3$.
\end{itemize}

\emph{Order-$3$ Symbol Generation}: Consider a random linear combination $\alpha_{1}u^{[1|1,2;3]}+\alpha_{2}u^{[1|1,3;2]}$. If we deliver this quantity to all three receivers, then
\begin{itemize}
\item RX$_1$ obtains a linear equation in terms of its own desired symbols,
\item since RX$_2$ has $u^{[1|1,3;2]}$, it can cancel $u^{[1|1,3;2]}$ to obtain $u^{[1|1,2;3]}$,
\item since RX$_3$ has $u^{[1|1,2;3]}$, it can cancel $u^{[1|1,2;3]}$ to obtain $u^{[1|1,3;2]}$.
\end{itemize}
Therefore, $\alpha_{1}u^{[1|1,2;3]}+\alpha_{2}u^{[1|1,3;2]}$ is desired by all three receivers. By similar arguments, one can conclude that $\beta_{1}u^{[2|2,1;3]}+\beta_{2}u^{[2|2,3;1]}$ and $\gamma_{1}u^{[3|1,3;2]}+\gamma_{2}u^{[3|2,3;1]}$ are desired by all three receivers, where $\beta_1$, $\beta_2$, $\gamma_1$, and $\gamma_2$ are random coefficients. According to our terminology, we define the following order-$3$ symbols.
\begin{align}
u^{[1|1,2,3]} &\Def \alpha_{1}u^{[1|1,2;3]}+\alpha_{2}u^{[1|1,3;2]}, \\
u^{[2|1,2,3]} &\Def \beta_{1}u^{[2|1,2;3]}+\beta_{2}u^{[2|2,3;1]}, \\
u^{[3|1,2,3]} &\Def \gamma_{1}u^{[3|1,3;2]}+\gamma_{2}u^{[3|2,3;1]}.
\end{align}

\emph{Order-$(1,2)$ Symbol Generation}: Although delivering $u^{[1|1,2,3]}$, $u^{[2|1,2,3]}$, and $u^{[3|1,2,3]}$ to all three receivers will provide each of them with useful information about its desired symbols as discussed above, it is not still sufficient to achieve the goals (a), (b), and (c). To be more specific, recall that RX$_1$ needs to obtain \emph{both} symbols $u^{[1|1,2;3]}$ and $u^{[1|1,3;2]}$. Thus, assuming $u^{[1|1,2,3]}$ has been delivered to all three receivers, RX$_1$ still needs an extra equation in terms of $u^{[1|1,2;3]}$ and $u^{[1|1,3;2]}$. To obtain this extra equation, we notice that by delivering $u^{[1|1,2,3]}$ to all three receivers, both RX$_2$ and RX$_3$ will have both symbols $u^{[1|1,2;3]}$ and $u^{[1|1,3;2]}$. Therefore, any random linear combination $\alpha'_{1}u^{[1|1,2;3]}+\alpha'_{2}u^{[1|1,3;2]}$ can be considered as the extra equation required at RX$_1$ which is also available at RX$_2$ and RX$_3$. Therefore, we can define the following $(1;2,3)$-symbol at TX$_1$.
\begin{equation}
u^{[1|1;2,3]} \Def \alpha'_{1}u^{[1|1,2;3]}+\alpha'_{2}u^{[1|1,3;2]}.
\end{equation}
Using the same argument for RX$_2$ and RX$_3$, the $(2;1,3)$-symbol and  $(3;1,2)$-symbol
\begin{align}
u^{[2|2;1,3]} &\Def \beta'_{1}u^{[2|1,2;3]}+\beta'_{2}u^{[2|2,3;1]}, \\
u^{[3|3;1,2]} &\Def \gamma'_{1}u^{[3|1,3;2]}+\gamma'_{2}u^{[3|2,3;1]},
\end{align}
can be defined, where $\beta'_1$, $\beta'_2$, $\gamma'_1$, and $\gamma'_2$ are random coefficients. 

To summarize, one can achieve the goals (a), (b), and (c) if
\begin{itemize}
\item[I.] $u^{[1|1,2,3]}$, $u^{[2|1,2,3]}$, and $u^{[3|1,2,3]}$ are delivered to all three receivers.
\item[II.] $u^{[1|1;2,3]}$, $u^{[2|2;1,3]}$, and $u^{[3|3;1,2]}$ are respectively delivered to RX$_1$, RX$_2$, and RX$_3$.
\end{itemize}
The goals I and II will be accomplished in the next phase.

\vspace{3mm}
$\bullet$  \underline{\textit{Phase $3$-I ($3$-user IC)}:}
\vspace{3mm}

In this subphase, which takes three time slots, we fulfill the goal I as follows: Using time division in three consecutive time slots, the three symbols $u^{[1|1,2,3]}$, $u^{[2|1,2,3]}$, and $u^{[3|1,2,3]}$ will be delivered to all three receivers.

\vspace{3mm}
$\bullet$ \underline{\textit{Phase $3$-II ($3$-user IC)}:}
\vspace{3mm}

In this subphase, the goal II is accomplished in one time slot by simultaneous transmission of $u^{[1|1;2,3]}$, $u^{[2|2;1,3]}$, and $u^{[3|3;1,2]}$ by TX$_1$, TX$_2$, and TX$_3$, respectively.

Finally, in order to compute the achieved DoF, we note that a total of $3\times 12=36$ fresh information symbols were fed to the channel in phase $1$. To deliver these information symbols to their intended receivers, we spent $3\times5=15$ time slots in phase $1$, $3\times4=12$ time slots in phase $2$, three time slots in subphase $3$-I, and one time slot in subphase $3$-II. Therefore, our achieved DoF is equal to
\begin{equation}
\DoFa_1^\text{IC}(3)=\frac{36}{15+12+3+1}=\frac{36}{31}.
\end{equation}

One finally notes that the proposed transmission scheme starting from the phase $2$ was dedicated to transmission of order-$2$ messages to the receivers. Therefore, we have proved that $\DoFa_2^\text{IC}(3)=\frac{18}{12+3+1}=\frac{9}{8}$ is achievable in the $3$-user IC with delayed CSIT as suggested by \cref{Eq:DoF_m(K)-IC}. Also, $\DoFa_3^\text{IC}(3)=1$ was trivially achieved using time division in the phase $3$-I.

\subsection{The $K$-user SISO Interference Channel}
\label{Sec:SISO-ICK}
\begin{figure*}
\centering
\includegraphics{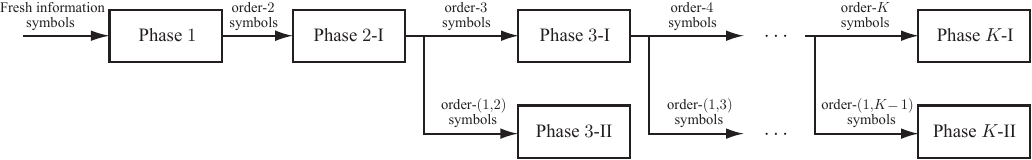}
\caption{Block diagram of the proposed multi-phase transmission scheme for the $K$-user IC, $K\geq 3$.}
\label{Fig:IC-BlockDiagram}
\end{figure*}

In this section, we elaborate on our transmission scheme for the $K$-user SISO IC with delayed CSIT and $K>3$. The transmission scheme is a multi-phase scheme wherein the fresh information symbols are fed to the system in phase $1$ towards generating order-$2$ symbols. The remaining phases are responsible for generating higher order symbols and finally providing each receiver with appropriate equations to resolve its own information symbols. \cref{Fig:IC-BlockDiagram} depicts a high-level block diagram for the proposed multi-phase scheme.

\vspace{3mm}
$\bullet$ \underline{\textit{Phase $1$ ($K$-user IC)}:}
\vspace{3mm}

\emph{Redundancy Transmission}: In this phase, each transmitter transmits $(K-1)^2+1$ random linear combinations of $(K-1)^2$ information symbols in $(K-1)^2+1$ time slots. Let ${\xu^{[i]}\Def [u_1^{[i]},u_2^{[i]},\cdots,u_{(K-1)^2}^{[i]}]^T}$ be the information vector of TX$_i$. For any $1\leq i\leq K$, define
\begin{align}
\xC^{[i]}\Def \left[\xc^{[i]}(1)|\xc^{[i]}(2)|\cdots|\xc^{[i]}({(K-1)^2+1})\right]^T,\nonumber
\end{align}
where $\xc^{[i]}(t)$ is the $(K-1)^2\times1$ vector of the random coefficients employed by TX$_i$ in time slot $t$, ${1\leq t \leq (K-1)^2+1}$. Then, ignoring the noise, RX$_j$ receives the vector
\begin{equation}
\xy_j=\xD_{j1}\xC^{[1]}\xu^{[1]}+\xD_{j2}\xC^{[2]}\xu^{[2]}+\cdots+\xD_{jK}\xC^{[K]}\xu^{[K]}, \nonumber
\end{equation}
of $(K-1)^2+1$ channel output symbols, where $\xD_{ji}$ is a diagonal matrix of size ${[(K-1)^2+1]\times[(K-1)^2+1]}$ containing the channel coefficients $h_{ji}(t)$, ${1\leq t \leq (K-1)^2+1}$, on its main diagonal.

\emph{PIN}: Since $\xD_{ji}$ and $\xC^{[i]}$ are full rank almost surely and independent of each other, their product is also full rank almost surely. Hence, defining $\xQ_{ji}\Def \xD_{ji}\xC^{[i]}$, ${1\leq i,j\leq K}$,  $\xQ_{ji}$ is a full rank matrix of size $[(K-1)^2+1]\times(K-1)^2$, and so, its left null space is one dimensional. Therefore, there exist nonzero vectors $\xomega_{ji}=[\omega_{ji1},\omega_{ji2},\cdots,\omega_{ji((K-1)^2+1)}]^T$ such that
\begin{equation}
\xQ^T_{ji} \xomega_{ji} = \mathbf{0}_{(K-1)^2\times1}, \qquad 1\leq i,j\leq K.
\end{equation}
Thus, for any $1\leq j\leq K$ and any $i\in \xxS_K \backslash \{j\}$, RX$_j$ can null out $\xu^{[i]}$ from its received vector and construct
\begin{align}
\xy^T_j\xomega_{ji}&=\sum_{i'\in \xxS_K \backslash \{i\}}(\xu^{[i']})^T\xQ_{ji'}^T\xomega_{ji}\nonumber \\
&=(\xu^{[j]})^T\xQ_{jj}^T\xomega_{ji}+\sum_{i'\in \xxS_K \backslash \{i,j\}}(\xu^{[i']})^T\xQ_{ji'}^T\xomega_{ji}. \label{Eq:yvect_Kuser}
\end{align}

\emph{Order-$2$ Symbol Generation}: We note that $(\xu^{[i']})^T\xQ_{ji'}^T\xomega_{ji}$, $i'\in \xxS_K \backslash \{i,j\}$, is an equation solely in terms of $\xu^{[i']}$, and thus, is desired by RX$_{i'}$. It is easy to see that if we deliver \emph{all} $K-2$ quantities $(\xu^{[i']})^T\xQ_{ji'}^T\xomega_{ji}$, $i'\in \xxS_K \backslash \{i,j\}$, to RX$_j$, then RX$_j$ can cancel their contributions from \cref{Eq:yvect_Kuser} to obtain $(\xu^{[j]})^T\xQ_{jj}^T\xomega_{ji}$, which is a desired equation for RX$_j$. Therefore, one can define $K-2$ order-2 $(i',j)$-symbols available at TX$_{i'}$ by
\begin{equation}
\label{Eq:order-2-IC}
u^{[i'|i',j]}\Def (\xu^{[i']})^T\xQ_{ji'}^T\xomega_{ji}, \qquad i'\in \xxS_K \backslash \{i,j\}.
\end{equation}

Since for a fixed $j$ there are $K-1$ choices of $i\in \xxS_K \backslash \{j\}$, a total of $(K-1)(K-2)$ order-$2$ symbols of the form $u^{[i|i,j]}$, $i\in \xxS_K \backslash \{j\}$, will be constructed for a fixed $j$. These symbols, if delivered, will provide RX$_j$ with $K-1$ equations solely in terms of $\xu^{[j]}$ while providing every RX$_i$, $i\in \xxS_K \backslash \{j\}$, with $K-2$ equations in terms of $\xu^{[i]}$.

Since there are $K$ choices for RX$_j$, $1\leq j\leq K$, a total of $K(K-1)(K-2)$ order-$2$ symbols $u^{[i|i,j]}$, $i\in \xxS_K \backslash \{j\}$, are generated by the end of phase $1$. After delivering all these symbols to their intended pairs of receivers, every receiver will be provided with $K-1+(K-1)(K-2)=(K-1)^2$ linear equations in terms of its own information symbols. Namely, RX$_j$ will obtain the following $(K-1)^2$ linear equations in terms of $\xu^{[j]}$.
\begin{align}
&(\xu^{[j]})^T \xQ_{jj}^T \xomega_{ji_1},\qquad i_1\in \xxS_K\backslash \{j\}, \label{Eq:Phase1-ICK1}\\
&(\xu^{[j]})^T \xQ_{i_2j}^T \xomega_{i_2i_3},\qquad i_2, i_3\in \xxS_K\backslash \{j\}, \hspace{3mm}i_2\neq i_3. \label{Eq:Phase1-ICK2}
\end{align}
It is proved in Appendix \ref{App:LinearInependence} that these $(K-1)^2$ linear combinations are linearly independent almost surely, and thus, each receiver can resolve all its $(K-1)^2$ information symbols. 

Finally, it takes $\frac{K(K-1)(K-2)}{\DoFa_2^\text{IC}(K)}$ time slots to deliver all the order-$2$ symbols generated in phase $1$ to their intended pairs of receivers. Hence, one can write
\begin{equation}
\label{Eq:DoF(K)-IC1}
\DoFa_1^\text{IC}(K)=\frac{(K-1)^2K}{(K-1)^2+1+\frac{K(K-1)(K-2)}{\DoFa_2^\text{IC}(K)}}.
\end{equation}

\vspace{3mm}
$\bullet$ \underline{\textit{Phase $m$-I, $2\leq m \leq K-1$ ($K$-user IC)}:}
\vspace{3mm}

This subphase takes a total of $N^\text{IC-I}_m$ order-$m$ symbols of the form $u^{[i|\xxS_m]}$, ${i\in \xxS_m}$, and transmits them over the channel in $T_m^\text{IC}$ time slots. Then, a total of $N^\text{IC-I}_{m+1}$ order-$(m+1)$ symbols of the form $u^{[i|\xxS_{m+1}]}$, ${i\in \xxS_{m+1}}$, together with $N^\text{IC-II}_{m+1}$ symbols of the form $u^{[i|i;\xxS_{m+1}\backslash \{i\}]}$, ${i\in \xxS_{m+1}}$, are generated such that if the generated symbols are delivered to their intended receiver(s), then every subset $\xxS_m$ of cardinality $m$ of receivers will be able to decode all the $\xxS_m$-symbols transmitted in this subphase. The parameters $N^\text{IC-I}_m$, $T_m^\text{IC}$, $N^\text{IC-I}_{m+1}$, and $N^\text{IC-II}_{m+1}$ are given by
\begin{align}
N^\text{IC-I}_m&=m[2(K-m)+1]\binom{K}{m}, \label{Eq:N_m-IC-I}\\
T^\text{IC}_m &= m(K-m+1)\binom{K}{m}, \label{Eq:T_m} \\
N^\text{IC-I}_{m+1}&=(m^2-1)\binom{K}{m+1}, \label{Eq:N_m+1-IC-I}\\
N^\text{IC-II}_{m+1}&=(m+1)\binom{K}{m+1}. \label{Eq:N_m+1-IC-II}
\end{align}
The details of transmission in this phase are as follows.

\emph{Redundancy Transmission}: Fix $\xxS_m \subset \xxS_K$ and sort the elements of $\xxS_m$ in ascending cyclic order. Fix $i_1\in \xxS_m$ and let $i_2\in \xxS_m$ be the element which comes immediately after $i_1$ in that ordering. Consider vector
\begin{align}
\xu^{[i_1|\xxS_m]}\Def \left[u_1^{[i_1|\xxS_m]},u_2^{[i_1|\xxS_m]},\cdots,u_{K-m+1}^{[i_1|\xxS_m]}\right]^T \nonumber
\end{align}
of $K-m+1$ $\xxS_m$-symbols available at TX$_{i_1}$ and vector
\begin{align}
\xu^{[i_2|\xxS_m]}\Def \left[u_1^{[i_2|\xxS_m]},u_2^{[i_2|\xxS_m]},\cdots,u_{K-m}^{[i_2|\xxS_m]}\right]^T \nonumber
\end{align}
 of $K-m$ $\xxS_m$-symbols available at TX$_{i_2}$. During the first $K-m+1$ time slots of this subphase, TX$_{i_1}$  and TX$_{i_2}$ transmit $K-m+1$ random linear combinations of elements of $\xu^{[i_1|\xxS_m]}$ and $\xu^{[i_2|\xxS_m]}$, respectively, while the rest of transmitters are silent. Let $\xc^{[i_1|\xxS_m]}(t)$ (resp.\ $\xc^{[i_2|\xxS_m]}(t)$) be the $(K-m+1)\times 1$ vector (resp.\ $(K-m)\times 1$  vector) of random coefficients employed by TX$_{i_1}$ (resp.\ TX$_{i_2}$) in time slot $t$, $1\leq t \leq K-m+1$. Then, ignoring the noise, by the end of these time slots, RX$_j$ receives the vector of $K-m+1$ channel output symbols
\begin{align}
\xy_j&= \xD_{ji_1}\xC^{[i_1|\xxS_m]}\xu^{[i_1|\xxS_m]}+\xD_{ji_2}\xC^{[i_2|\xxS_m]}\xu^{[i_2|\xxS_m]} \nonumber \\
&=\xQ_{ji_1}\xu^{[i_1|\xxS_m]}+\xQ_{ji_2}\xu^{[i_2|\xxS_m]},
\end{align}
where $\xC^{[i_1|\xxS_m]}$ and $\xC^{[i_2|\xxS_m]}$ are defined as
\begin{align}
\xC^{[i_1|\xxS_m]} &\Def  \left[\xc^{[i_1|\xxS_m]}(1)|\cdots|\xc^{[i_1|\xxS_m]}({K-m+1})\right]^T, \\
\xC^{[i_2|\xxS_m]} &\Def  \left[\xc^{[i_2|\xxS_m]}(1)|\cdots|\xc^{[i_2|\xxS_m]}({K-m+1})\right]^T, 
\end{align}
$\xD_{ji_1}$ and $\xD_{ji_2}$ are diagonal matrices of size $(K-m+1)\times(K-m+1)$ containing the channel coefficients $h_{ji_1}(t)$ and $h_{ji_2}(t)$, ${1\leq t \leq K-m+1}$, on their main diagonal, respectively, and $\xQ_{ji_1}$ and $\xQ_{ji_2}$ are defined as
\begin{align}
\xQ_{ji_1} &\Def \xD_{ji_1}\xC^{[i_1|\xxS_m]},\\
\xQ_{ji_2} &\Def \xD_{ji_2}\xC^{[i_2|\xxS_m]}.
\end{align}

Therefore, each receiver RX$_j$, $j\in \xxS_m$, obtains $K-m+1$ desired linearly independent equations in terms of the $2(K-m)+1$ transmitted $\xxS_m$-symbols, and thus, needs $K-m$ extra equations to resolve all the transmitted $\xxS_m$-symbols. 

\emph{PIN}: It is easily verified that $\xQ_{ji_2}$ is a full rank matrix of size $(K-m+1)\times(K-m)$ almost surely, and so, its left null space is one dimensional. Specifically, there exist nonzero vectors $\xomega_{j'i_2}$ such that
\begin{equation}
\xQ_{j'i_2}^T \xomega_{j'i_2}=\mathbf{0}, \qquad j' \in \xxS_K \backslash \xxS_m.
\end{equation}
Hence, each receiver RX$_{j'}$, $j' \in \xxS_K \backslash \xxS_m$, can null out $\xu^{[i_2|\xxS_m]}$ from its received vector and construct
\begin{align}
\xy^T_{j'} \xomega_{j'i_2}=(\xu^{[i_1|\xxS_m]})^T\xQ^T_{j'i_1}\xomega_{j'i_2},
\end{align}
which is a linear combination in terms of $\xu^{[i_1|\xxS_m]}$, and thus, if delivered to all receivers RX$_j$, $j\in \xxS_m$, can provide each of them with an extra equation in terms of their desired $\xxS_m$-symbols. On the other hand, the above linear combination is solely in terms of $\xu^{[i_1|\xxS_m]}$ (available at TX$_{i_1}$) and the channel coefficients (available at TX$_{i_1}$, due to the delayed CSIT assumption, by the end of these $K-m+1$ time slots). Therefore, based on our terminology, one can define
\begin{equation}
\label{Eq:SideInf-ICm}
u^{[i_1|\xxS_m;j']} \Def  (\xu^{[i_1|\xxS_m]})^T\xQ^T_{j'i_1}\xomega_{j'i_2}, \qquad j'\in \xxS_K \backslash \xxS_m.
\end{equation}
After delivering all these side information symbols to all receivers RX$_j$, $j\in \xxS_m$, each of them will obtain $2(K-m)+1$ linear combinations in terms of the $2(K-m)+1$ transmitted $\xxS_m$-symbols. Namely, RX$_j$, $j\in \xxS_m$, will obtain the linear combinations
\begin{align}
&\xQ_{ji_1}\xu^{[i_1|\xxS_m]}+\xQ_{ji_2}\xu^{[i_2|\xxS_m]} \label{Eq:Phasem-ICK1}\\
 &(\xu^{[i_1|\xxS_m]})^T\xQ^T_{j'i_1}\xomega_{j'i_2}, \qquad j'\in \xxS_K \backslash \xxS_m. \label{Eq:Phasem-ICK2}
\end{align}
which are shown to be linearly independent almost surely in Appendix \ref{App:LinearInependence-S_m}. This enables RX$_j$ to solve them for $\xu^{[i_1|\xxS_m]}$ and $\xu^{[i_2|\xxS_m]}$.

We repeat the same procedure for every choice of $i_1\in \xxS_m$, \ie for each choice, we spend $K-m+1$ time slots to transmit $2(K-m)+1$ $\xxS_m$-symbols and generate $K-m$ side information symbols. This implies transmission of a total of $m[2(K-m)+1]$ $\xxS_m$-symbols in $m(K-m+1)$ time slots and generation of $m(K-m)$ side information symbols. Since $\xxS_m\subset \xxS_K$ could be any subset with cardinality $m$, we transmit a total of $N^\text{IC-I}_m$ order-$m$ symbols in $T^\text{IC}_m$ time slots and generate $m(K-m)\binom{K}{m}$ side information symbols, where $N^\text{IC-I}_m$ and $T^\text{IC}_m$ are given by \cref{Eq:N_m-IC-I,Eq:T_m}.

\emph{Order-$(m+1)$ Symbol Generation}: Fix a subset $\xxS_{m+1}\subset \xxS_K$ and an index $i_1\in \xxS_{m+1}$. Since for any $j' \in \xxS_{m+1} \backslash \{i_1\}$ we have generated exactly one side information symbol $u^{[i_1|\xxS_{m+1} \backslash \{j'\};j']}$, there exist $m$ symbols $u^{[i_1|\xxS_{m+1} \backslash \{j'\};j']}$, $j' \in \xxS_{m+1} \backslash \{i_1\}$, for a fixed $\xxS_{m+1}$ and a fixed $i_1\in \xxS_{m+1}$. Moreover, every receiver RX$_{j'}$, $j'\in \xxS_{m+1}\backslash \{i_1\}$, has exactly one of these $m$ symbols and wishes to obtain the rest, while RX$_{i_1}$ wishes to obtain all the $m$ symbols. Therefore, if we deliver $m-1$ random linear combinations of these $m$ symbols to all receivers in $\xxS_{m+1}$, then each of them (except for RX$_{i_1}$) will remove its known side information and obtain $m-1$ linearly independent equations in terms of the $m-1$ desired symbols, and hence, decode all desired symbols. Thus, these $m-1$ random linear combinations are defined as $m-1$ $\xxS_{m+1}$-symbols $u_\ell^{[i_1|\xxS_{m+1}]}$, $1\leq \ell \leq m-1$.

\emph{Order-$(1,m)$ Symbol Generation}: Since RX$_{i_1}$ wishes to obtain all the $m$ symbols $u^{[i_1|\xxS_{m+1} \backslash \{j'\};j']}$, $j'\in \xxS_{m+1}\backslash \{i_1\}$, after delivering the above $m-1$ random linear combinations to RX$_{i_1}$, it still requires one extra linearly independent equation to resolve all its desired symbols. On the other hand, recall that after delivering all the $\xxS_{m+1}$-symbols defined above to all receivers RX$_{j'}$, $j'\in\xxS_{m+1}$, every receiver RX$_{j'}$, $j'\in \xxS_{m+1} \backslash \{i_1\}$, will be able to obtain \emph{all} the $m$ symbols $u^{[i_1|\xxS_{m+1}\backslash \{j'\};j']}$, $j'\in \xxS_{m+1} \backslash \{i_1\}$. Thereafter, any linear combination of the symbols $u^{[i_1|\xxS_{m+1}\backslash \{j'\};j']}$, $j'\in \xxS_{m+1} \backslash \{i_1\}$, will be available at every receiver RX$_{j'}$, $j'\in \xxS_{m+1} \backslash \{i_1\}$. Specifically, the extra random linear combination of $u^{[i_1|\xxS_{m+1}\backslash \{j'\};j']}$, $j'\in \xxS_{m+1}\backslash \{i_1\}$ which is required by RX$_{i_1}$ can be denoted as $u^{[i_1|i_1;\xxS_{m+1}\backslash \{i_1\}]}$.

In summary, since there are $\binom{K}{m+1}$ choices of $\xxS_{m+1}\subset \xxS_K$, and $m+1$ choices of $i_1\in \xxS_{m+1}$ for each $\xxS_{m+1}$, a total of $N^\text{IC-I}_{m+1}$ order-$(m+1)$ $\xxS_{m+1}$-symbols and $N^\text{IC-II}_{m+1}$ order-$(1,m)$ $(i_1;\xxS_{m+1}\backslash \{i_1\})$-symbols will be generated where $N^\text{IC-I}_{m+1}$ and $N^\text{IC-II}_{m+1}$ are given by \cref{Eq:N_m+1-IC-I,Eq:N_m+1-IC-II}. If we deliver all the $\xxS_{m+1}$-symbols and $(i_1;\xxS_{m+1}\backslash \{i_1\})$-symbols, $\xxS_{m+1}\subset \xxS_K$, $i_1\in \xxS_{m+1}$, to their intended receiver(s), then each receiver will be able to decode all its desired order-$m$ symbols transmitted in this subphase. This will be accomplished during the next phases.

\vspace{3mm}
$\bullet$ \underline{\textit{Phase $m$-II, $3\leq m \leq K$ ($K$-user IC)}:}
\vspace{3mm}

In this subphase, each time slot is dedicated to transmission of the order-$(1,m-1)$ symbols $u^{[i|i;\xxS_m\backslash \{i\}]}$, $i\in \xxS_m$, for a fixed $\xxS_m$, $\xxS_{m}\subset \xxS_K$. In particular, during the time slot dedicated to $\xxS_m$, every transmitter TX$_i$, $i\in \xxS_m$, transmits $u^{[i|i;\xxS_m\backslash \{i\}]}$, simultaneously. Since each receiver RX$_j$, $j\in \xxS_m$, has all symbols $u^{[i|i;\xxS_m\backslash \{i\}]}$, $i\in \xxS_m \backslash \{j\}$, it will decode its desired symbol (\ie $u^{[j|j;\xxS_m\backslash \{j\}]}$) after this time slot. If we denote by $\DoFa^\text{IC-II}_m(K)$ the achievable DoF of transmitting $(i;\xxS_m\backslash \{i\})$-symbols over the $K$-user SISO IC with delayed CSIT, one can write
\begin{equation}
\label{Eq:DoF_m(K)-IC-II}
\DoFa^\text{IC-II}_m(K)= m, \qquad 3\leq m \leq K.
\end{equation}

\vspace{3mm}
$\bullet$ \underline{\textit{Phase $K$-I ($K$-user IC)}:}
\vspace{3mm}

In this subphase, during each time slot, an order-$K$ symbol $u^{[i|\xxS_K]}$, $i\in \xxS_K$, is transmitted by TX$_i$ while the other transmitters are silent. After each time slot, ignoring the noise, each receiver receives the transmitted symbol without any interference. This implies that 
\begin{equation}
\label{Eq:DoF_K(K)-IC1}
\DoFa_K^\text{IC}(K)=1.
\end{equation}

Combining \cref{Eq:N_m-IC-I,Eq:T_m,Eq:N_m+1-IC-I,Eq:N_m+1-IC-II,Eq:DoF_m(K)-IC-II}, we conclude that for $2\leq m\leq K-1$,
\begin{align}
\DoFa_m^\text{IC}&(K)=\frac{N^\text{IC-I}_m}{T^\text{IC}_m+\frac{N^\text{IC-II}_{m+1}}{\DoFa^\text{IC-II}_{m+1}(K)}+\frac{N^\text{IC-I}_{m+1}}{\DoFa_{m+1}^\text{IC}(K)}} \nonumber \\
&=\frac{m[2(K-m)+1]\binom{K}{m}}{m(K-m+1)\binom{K}{m}+\frac{(m+1)\binom{K}{m+1}}{m+1}+\frac{(m^2-1)\binom{K}{m+1}}{\DoFa_{m+1}^\text{IC}(K)}} \nonumber \\
&= \frac{m[2(K-m)+1]}{m(K-m+1)+\frac{K-m}{m+1}+\frac{(m-1)(K-m)}{\DoFa_{m+1}^\text{IC}(K)}}. \label{Eq:DoF_m(K)-IC1}% \nonumber \\
%&\hspace{45mm}2\leq m\leq K-1. \label{Eq:DoF_m(K)-IC1}
\end{align}

It is shown in Appendix \ref{App:DoF-IC} that  \cref{Eq:DoF_m(K)-IC} is a closed form solution to the recursive equation \cref{Eq:DoF_m(K)-IC1} with the initial condition \eqref{Eq:DoF_K(K)-IC1} and $2\leq m \leq K$. As a result, for $m=2$, it is shown that
\begin{equation}
\label{Eq:DoF_2(K)}
\DoFa^\text{IC}_2(K)= \frac{1}{1-A_2(K)},
\end{equation}
where $A_2(K)$ is given in \cref{Eq:A_2(K)}. \Cref{Eq:DoF(K)-IC} immediately follows from \cref{Eq:DoF(K)-IC1,Eq:DoF_2(K),Eq:A_2(K)}.

\section{Proof of Theorem 2}
\label{Sec:SISO-X}
For $K=2$, our transmission scheme reduces to a modified version of the scheme proposed in \cite{Ghasemi2011Xchannel} and achieves the same DoF of $\frac{6}{5}$. Hence, we would rather start with $K=3$ and elaborate on our transmission scheme for the $2\times 3$ X channel with delayed CSIT. We show that it achieves $\DoFa_1^\text{X}(2,3)=\frac{9}{7}$ and $\DoFa_2^\text{X}(2,3)=\frac{9}{8}$, as suggested by \cref{Eq:DoF(K)-X,Eq:DoF_m(K)-X}. Finally, we will proceed with the general $2\times K$ case.

\subsection{The $2\times 3$ SISO X Channel}

In this section, we prove that $\DoFa_1^\text{X}(2,3)=\frac{9}{7}$ and $\DoFa_2^\text{X}(2,3)=\frac{9}{8}$ are achievable in the $2\times3$ SISO X channel with delayed CSIT which is depicted in \cref{Fig:X2-3}. To this end, we propose a transmission scheme which has three distinct phases:
\begin{figure}
\centering
\includegraphics[scale=1]{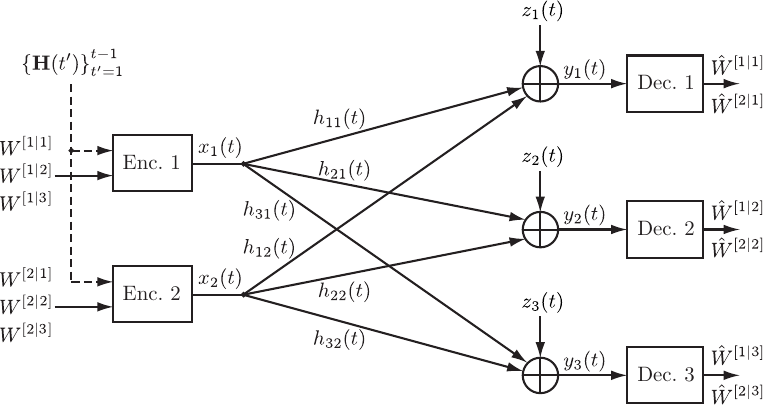}
\caption{The $2\times3$ SISO X channel with delayed CSIT.}
\label{Fig:X2-3}
\end{figure}

\vspace{3mm}
$\bullet$ \underline{\textit{Phase $1$ ($2\times 3$ X Channel)}:}
\vspace{3mm}

This phase takes $9$ time slots to transmit $15$ information symbols as follows. 

\emph{Redundancy Transmission}: Fix ${i_1=1}$ and ${i_2=2}$. During the first $3$ time slots, $5$ information symbols $\xu^{[i_1|1]}\Def[u_1^{[i_1|1]}, u_2^{[i_1|1]}, u_3^{[i_1|1]}]^T$ and ${\xu^{[i_2|1]}\Def[u_1^{[i_2|1]}, u_2^{[i_2|1]}]^T}$ (all intended for RX$_1$) are transmitted by TX$_{i_1}$ and TX$_{i_2}$, respectively. In particular, in each of these $3$ time slots, TX$_{i_1}$ transmits a random linear combination of $u_1^{[i_1|1]}$, $u_2^{[i_1|1]}$, and $u_3^{[i_1|1]}$ while TX$_{i_2}$ transmits a random linear combination of $u_1^{[i_2|1]}$ and $u_2^{[i_2|1]}$. 

Denote by
\begin{align}
\xc^{[i_1|1]}(t)\Def\left[c^{[i_1|1]}_1(t),c^{[i_1|1]}_2(t),c^{[i_1|1]}_3(t)\right]^T
\end{align}
and 
\begin{align}
\xc^{[i_2|1]}(t)\Def\left[c^{[i_2|1]}_1(t),c^{[i_2|1]}_2(t)\right]^T
\end{align}
the vectors containing random coefficients of the linear combinations transmitted by TX$_{i_1}$ and TX$_{i_2}$, respectively, over time slot $t$, $1\leq t \leq 3$. By the end of these $3$ time slots, every receiver obtains $3$ linearly independent equations in terms of the $5$ transmitted information symbols almost surely. Thus, RX$_1$ requires \emph{two} more linearly independent equations to resolve its $5$ desired information symbols. Now, consider the linear combinations received RX$_2$ and RX$_3$ during time slot $t$, $1\leq t \leq 3$, \ie
\begin{align}
y_j(t)&=\sum_{k=1}^2h_{ji_k}(t)x_{i_k}(t) \nonumber \\
&=\sum_{k=1}^2h_{ji_k}(t)\left(\xc^{[i_k|1]}(t)\right)^T\xu^{[i_k|1]}, \quad j=2,3.
\end{align}
In a vector form, one can write
\begin{equation}
\label{Eq:y_vector-X}
\xy_{j|1}=\sum_{k=1}^2\xD_{ji_k|1}\xC^{[i_k|1]}\xu^{[i_k|1]}, \quad j=2,3,
\end{equation}
where $\xy_{j|1}$ is the vector of $3$ received symbols at RX$_j$ during these $3$ time slots, $\xD_{ji_k|1}$ is the $3\times3$ diagonal matrix containing $h_{ji_k}(t)$, $1\leq t \leq 3$, on its main diagonal, and $\xC^{[i_1|1]}$ (resp.\ $\xC^{[i_2|1]}$) is the $3\times3$ (resp.\ $3\times2$) matrix containing the random coefficients employed by TX$_{i_1}$ (resp.\ TX$_{i_2}$) during these $3$ time slots, i.e.,
\begin{align}
\xC^{[i_k|1]} &{\Def} {\left[\xc^{[i_k|1]}(1)|\xc^{[i_k|1]}(2)|\xc^{[i_k|1]}(3)\right]^T}, \quad k=1,2.
\end{align}

\emph{PIN}: Since the elements of $\xC^{[i_1|1]}$ and $\xC^{[i_2|1]}$ are i.i.d., they are full rank almost surely. Also, $\xD_{ji_k|1}$ is full rank almost surely and is independent of $\xC^{[i_k|1]}$. Therefore, $\xQ_{ji_k|1}\Def \xD_{ji_k|1}\xC^{[i_k|1]}$ is full rank almost surely. Specifically, $\xQ_{ji_2|1}$ is a full rank $3\times2$ matrix, and thus, its left null space is one dimensional almost surely. Let the $3\times1$ vector $\xomega_{ji_2|1}$ be in the left null space of $\xQ_{ji_2|1}$, i.e.,
\begin{equation}
\label{Eq:omega-X}
\xQ_{ji_2|1}^T\xomega_{ji_2|1}=\mathbf{0}_{2\times1}, \quad j=2,3.
\end{equation}

Using \cref{Eq:y_vector-X,Eq:omega-X}, RX$_j$, $j=2,3$, can null out $\xu^{[i_2|1]}$ from its received vector and obtain
\begin{align}
\xy_{j|1}^T\xomega_{ji_2|1}&=(\xu^{[i_1|1]})^T\xQ_{ji_1|1}^T\xomega_{ji_2|1}+(\xu^{[i_2|1]})^T\underbrace{\xQ_{ji_2|1}^T\xomega_{ji_2|1}}_\mathbf{0} \nonumber \\
&=(\xu^{[i_1|1]})^T\xQ_{ji_1|1}^T\xomega_{ji_2|1},
\end{align}
which is an equation solely in terms of $\xu^{[i_1|1]}$. Therefore, if we deliver $(\xu^{[i_1|1]})^T\xQ_{ji_1|1}^T\xomega_{ji_2|1}$, $j=2,3$, to RX$_1$, it will have enough equations to resolve its $5$ desired information symbols (it can be easily shown that these equations are linearly independent almost surely). Hence, two symbols $u^{[i_1|1;2]}$ and $u^{[i_1|1;3]}$ can be defined as
\begin{align}
\label{Eq:Side1-X}
u^{[i_1|1;j]}\Def(\xu^{[i_1|1]})^T\xQ_{ji_1|1}^T\xomega_{ji_2|1}, \quad j=2,3.
\end{align}

In the same way, the following $5$ fresh information symbols (now, all intended for RX$_2$) are transmitted during the next $3$ time slots
\begin{align}
\xu^{[i_1|2]}&\Def[u_1^{[i_1|2]}, u_2^{[i_1|2]}, u_3^{[i_1|2]}]^T,\\
\xu^{[i_2|2]}&\Def[u_1^{[i_2|2]}, u_2^{[i_2|2]}]^T,
\end{align}
and the following two side information symbols are generated
\begin{align}
\label{Eq:Side2-X}
u^{[i_1|2;j]}\Def(\xu^{[i_1|2]})^T\xQ_{ji_1|2}^T\xomega_{ji_2|2}, \quad j=1,3,
\end{align}
where $\xQ_{ji_1|2}^T$ and $\xomega_{ji_2|2}$ are similarly defined.

The same procedure is followed during the last $3$ time slots to transmit another $5$ fresh information symbols
\begin{align}
\xu^{[i_1|3]}&\Def[u_1^{[i_1|3]}, u_2^{[i_1|3]}, u_3^{[i_1|3]}]^T,\\
\xu^{[i_2|3]}&\Def[u_1^{[i_2|3]}, u_2^{[i_2|3]}]^T,
\end{align}
which are all intended for RX$_3$, and generate two side information symbols
\begin{align}
\label{Eq:Side3-X}
u^{[i_1|3;j]}\Def(\xu^{[i_1|3]})^T\xQ_{ji_1|3}^T\xomega_{ji_2|3}, \quad j=1,2,
\end{align}
with similar definitions of $\xQ_{ji_1|3}^T$ and $\xomega_{ji_2|3}$.

After these $9$ time slots, if we deliver the side information symbols defined in \cref{Eq:Side1-X,Eq:Side2-X,Eq:Side3-X} to their respective receivers, then each receiver will be able to decode all its own $5$ information symbols. 

\emph{Order-$2$ Symbol Generation}: Consider the linear combination $u^{[i_1|1;2]}+u^{[i_1|2;1]}$. If we deliver this linear combination to \emph{both} RX$_1$ and RX$_2$, then RX$_1$ can cancel $u^{[i_1|2;1]}$ to obtain $u^{[i_1|1;2]}$.  Similarly, RX$_2$ can cancel $u^{[i_1|1;2]}$ to obtain $u^{[i_1|2;1]}$. Note also that both $u^{[i_1|1;2]}$ and $u^{[i_1|2;1]}$ are available at TX$_{i_1}$, and so is their summation. Therefore, one can define the order-$2$ symbol
\begin{equation}
\label{Eq:Order2-symbol-X}
u^{[i_1|1,2]}\Def u^{[i_1|1;2]}+u^{[i_1|2;1]}
\end{equation}
which is available at TX$_{i_1}$. The following order-$2$ symbols can be similarly defined.
\begin{align}
u^{[i_1|1,3]}&\Def u^{[i_1|1;3]}+u^{[i_1|3;1]},\label{Eq:(1,3)-symbol-X}\\
u^{[i_1|2,3]}&\Def u^{[i_1|2;3]}+u^{[i_1|3;2]}\label{Eq:(2,3)-symbol-X}.
\end{align}
Our goal in phase $2$ is to deliver the above three order-$2$ symbols to their respective pairs of receivers.

\vspace{3mm}
$\bullet$ \underline{\textit{Phase $2$ ($2\times 3$ X Channel)}:}
\vspace{3mm}

This phase takes $12$ time slots to transmit $18$ order-$2$ symbols. Recall that in phase $1$ we generated only three order-$2$ symbols $u^{[i_1|1,2]}$, $u^{[i_1|1,3]}$, and $u^{[i_1|2,3]}$ which are all available at TX$_{i_1}$, where ${i_1=1}$. As we will see later, the following $18$ order-$2$ symbols are required for phase $2$.
\begin{equation}
u_k^{[i|1,2]}, u_k^{[i|1,3]}, u_k^{[i|2,3]}, \quad i=1,2, \quad1\leq k \leq 3.
\end{equation}
Therefore, we repeat phase $1$ three times with ${(i_1,i_2)=(1,2)}$ and three times with ${(i_1,i_2)=(2,1)}$ to generate the above $18$ order-$2$ symbols. The transmission in phase $2$ is then accomplished as follows.

\emph{Redundancy Transmission and PIN}: The first $4$ time slots of phase $2$ are dedicated to transmission of $6$ $(1,2)$-symbols $\{u_k^{[1|1,2]}\}_{k=1}^3$ and $\{u_k^{[2|1,2]}\}_{k=1}^3$. This is accomplished in exactly the same way as the first $4$ time slots of phase $2$ of the scheme proposed for the $3$-user IC in \cref{Sec:SISO-IC3}, and the side information symbols $u^{[1|1,2;3]}$ and $u^{[2|1,2;3]}$ will be generated at RX$_3$. 
Similar to phase $2$ of \cref{Sec:SISO-IC3}, the next $8$ time slots are dedicated to transmission of $6$ $(1,3)$-symbols and $6$ $(2,3)$-symbols. However, in contrast to \cref{Sec:SISO-IC3}, the $(1,3)$-symbols and $(2,3)$-symbols are here transmitted by TX$_1$ and TX$_2$. Hence, after these $8$ time slots, the side information $u^{[1|2,3;1]}$ and $u^{[2|2,3;1]}$ will be generated at RX$_1$ and the side information $u^{[1|1,3;2]}$ and $u^{[2|1,3;2]}$ will be generated at RX$_2$.

Therefore, after these $12$ time slots, our goal is reduced to
\begin{itemize}
\item[(a)] delivering $u^{[1|1,2;3]}$ and $u^{[2|1,2;3]}$ to both RX$_1$ and RX$_2$,
\item[(b)] delivering $u^{[1|1,3;2]}$ and $u^{[2|1,3;2]}$ to both RX$_1$ and RX$_3$,
\item[(c)] delivering $u^{[1|2,3;1]}$ and $u^{[2|2,3;1]}$ to both RX$_2$ and RX$_3$.
\end{itemize}

\emph{Order-$3$ Symbol Generation}: Now, consider $u^{[1|1,2;3]}$, $u^{[1|1,3;2]}$, and $u^{[1|2,3;1]}$. Note that these three symbols are available at TX$_1$, and so is any linear combination of them. Another observation is that each receiver has exactly one symbol out of these three symbols and requires the other \emph{two}. Hence, if we deliver \emph{two} random linear combinations of these three symbols to all receivers, then RX$_1$ can remove $u^{[1|2,3;1]}$ from them to obtain two random linear combinations solely in terms of $u^{[1|1,2;3]}$ and $u^{[1|1,3;2]}$, and so, solve them for $u^{[1|1,2;3]}$ and $u^{[1|1,3;2]}$. Likewise, RX$_2$ (resp.\ RX$_3$) can remove $u^{[1|1,3;2]}$ (resp.\ $u^{[1|1,2;3]}$) from the two random linear combinations and obtain two random linear equations solely in terms of its own pair of desired symbols, and resolve its desired symbols. Thus, the following two random linear combinations can be considered as order-$3$ symbols to be delivered to all three receivers in the next phase.
\begin{align}
u_1^{[1|1,2,3]}&\Def \alpha_1u^{[1|1,2;3]}+\alpha_2u^{[1|1,3;2]}+\alpha_3u^{[1|2,3;1]}, \\
u_2^{[1|1,2,3]}&\Def \alpha'_1u^{[1|1,2;3]}+\alpha'_2u^{[1|1,3;2]}+\alpha'_3u^{[1|2,3;1]}.
\end{align}
Using the same arguments, one can define the order-$3$ symbols
\begin{align}
u_1^{[2|1,2,3]}&\Def \beta_1u^{[2|1,2;3]}+\beta_2u^{[2|1,3;2]}+\beta_3u^{[2|2,3;1]}, \\
u_2^{[2|1,2,3]}&\Def \beta'_1u^{[2|1,2;3]}+\beta'_2u^{[2|1,3;2]}+\beta'_3u^{[2|2,3;1]},
\end{align}
where $\beta_i$ and $\beta'_i$, $1\leq i \leq 3$, are random coefficients.

\vspace{3mm}
$\bullet$ \underline{\textit{Phase $3$ ($2\times 3$ X Channel)}:}
\vspace{3mm}

Using time division in $4$ time slots, the $4$ order-$3$ symbols $u_1^{[1|1,2,3]}$, $u_2^{[1|1,2,3]}$, $u_1^{[2|1,2,3]}$, and $u_2^{[2|1,2,3]}$ will be delivered to all three receivers.

At the end, since we fed a total of $6\times15=90$ fresh information symbols to the system during $6\times9=54$ time slots in phase $1$, and spent $12$ time slots in phase $2$ and $4$ time slots in phase $3$, the achieved DoF is equal to
\begin{equation}
\DoFa_1^\text{X}(2,3)=\frac{90}{54+12+4}=\frac{9}{7}.
\end{equation}
Also, in view of phases $2$ and $3$, we have $\DoFa_2^\text{X}(2,3)=\frac{18}{12+4}=\frac{9}{8}$, and $\DoFa_3^\text{X}(2,3)=1$.

\subsection{The $2\times K$ SISO X Channel}
\begin{figure*}
\centering
\includegraphics{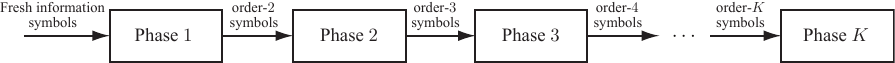}
\caption{Block diagram of the proposed multi-phase transmission scheme for the $2\times K$ X channel, $K\geq 2$.}
\label{Fig:X-BlockDiagram}
\end{figure*}

Our transmission scheme for the $2\times K$ SISO X channel with delayed CSIT is a multi-phase scheme as illustrated in \cref{Fig:X-BlockDiagram}. In particular, for every $m$, $1\leq m \leq K-1$, phase $m$ takes $N^\text{X}_m$ order-$m$ symbols of the form $u^{[i|\xxS_m]}$, ${i\in \{1,2\}}$, and transmits them over the channel in $T^\text{X}_m$ time slots. Then, a total of $N^\text{X}_{m+1}$ order-$(m+1)$ symbols of the form $u^{[i|\xxS_{m+1}]}$, ${i\in \{1,2\}}$, are generated such that if the generated symbols are delivered to their intended receivers, every subset $\xxS_m$ of cardinality $m$ of receivers will be able to decode all the $\xxS_m$-symbols transmitted in phase $m$. The parameters $N^\text{X}_m$, $T_m^\text{X}$, and $N^\text{X}_{m+1}$ are given by
\begin{align}
N^\text{X}_m&=2[2(K-m)+1]\binom{K}{m}, \label{Eq:N'_m}\\
T^\text{X}_m &= 2(K-m+1)\binom{K}{m}, \label{Eq:T'_m} \\
N^\text{X}_{m+1}&=2m\binom{K}{m+1}. \label{Eq:N'_m+1}
\end{align}
The following is a detailed description of phase $m$.

\vspace{3mm}
$\bullet$ \underline{\textit{Phase $m$, $1\leq m \leq K-1$ ($2\times K$ X Channel)}:}
\vspace{3mm}

\emph{Redundancy Transmission and PIN}: Fix $i_1=1$ and $i_2=2$. For every $\xxS_m \subset \xxS_K$, consider two vectors of $\xxS_m$-symbols
\begin{align}
\xu^{[i_1|\xxS_m]}&\Def \left[u_1^{[i_1|\xxS_m]},u_2^{[i_1|\xxS_m]},\cdots,u_{K-m+1}^{[i_1|\xxS_m]}\right]^T, \label{Eq:Order-m-X-i1}\\
\xu^{[i_2|\xxS_m]}&\Def \left[u_1^{[i_2|\xxS_m]},u_2^{[i_2|\xxS_m]},\cdots,u_{K-m}^{[i_2|\xxS_m]}\right]^T \label{Eq:Order-m-X-i2},
\end{align}
and transmit them exactly as in phase $m$-I of \cref{Sec:SISO-ICK}. More specifically, during $K-m+1$ time slots, TX$_{i_1}$  and TX$_{i_2}$ transmit $K-m+1$ random linear combinations of elements of $\xu^{[i_1|\xxS_m]}$ and $\xu^{[i_2|\xxS_m]}$, respectively. Using the same arguments as in phase $m$-I of \cref{Sec:SISO-ICK}, $K-m$ side information symbols of the form $u^{[i_1|\xxS_m;j']}$, $j'\in \xxS_K \backslash \xxS_m$, are generated after these $K-m+1$ time slots (see \cref{Eq:SideInf-ICm}). If we deliver all symbols $u^{[i_1|\xxS_m;j']}$, $j'\in \xxS_K \backslash \xxS_m$, to all receivers in $j\in \xxS_m$, then all of them will obtain enough linearly independent equations to decode all the $\xxS_m$-symbols in $\xu^{[i_1|\xxS_m]}$ and $\xu^{[i_2|\xxS_m]}$.

Therefore, for every $\xxS_m \subset \xxS_K$, a total of $2(K-m)+1$ $\xxS_m$-symbols are transmitted in $K-m+1$ time slots, and $K-m$ side information symbols are generated. Since there are $\binom{K}{m}$ choices of $\xxS_m \subset \xxS_K$, this implies transmission of $[2(K-m)+1]\binom{K}{m}$ order-$m$ symbols in $(K-m+1)\binom{K}{m}$ time slots and generation of $(K-m)\binom{K}{m}$ side information symbols. 

\emph{Order-$(m+1)$ Symbol Generation}: Fix a subset $\xxS_{m+1}\subset \xxS_K$. Since for every $j'\in \xxS_{m+1}$ we have generated exactly one side information symbol $u^{[i_1|\xxS_{m+1}\backslash \{j'\};j']}$, there exist $m+1$ symbols $u^{[i_1|\xxS_{m+1} \backslash \{j'\};j']}$, $j'\in \xxS_{m+1}$, for a fixed $\xxS_{m+1}$. Moreover, every receiver RX$_{j'}$, $j'\in \xxS_{m+1}$, has exactly one of these $m+1$ symbols and wishes to obtain the rest. Therefore, if we deliver $m$ random linear combinations of these $m+1$ symbols to all receivers in $\xxS_{m+1}$, each of them will remove its known side information and obtain $m$ linearly independent equations in terms of the $m$ desired symbols, and hence, decode all desired symbols. Thus, these $m$ random linear combinations are defined as $m$ $\xxS_{m+1}$-symbols $u_\ell^{[i_1|\xxS_{m+1}]}$, ${1\leq \ell \leq m}$. Since there are $\binom{K}{m+1}$ choices of $\xxS_{m+1}$, $\xxS_{m+1}\subset \xxS_K$, a total of $m\binom{K}{m+1}$ order-$(m+1)$ symbols will be generated as above.

Finally we note that, so far, we have only generated order-$(m+1)$ symbols of the form $u^{[i_1|\xxS_{m+1}]}$, with $i_1=1$, which are all available at TX$_{1}$. However, in order for phase $m+1$ to work, we need order-$(m+1)$ symbols of \emph{both} forms $u^{[1|\xxS_{m+1}]}$ and $u^{[2|\xxS_{m+1}]}$. This can be seen from \cref{Eq:Order-m-X-i1,Eq:Order-m-X-i2}. Therefore, we simply repeat phase $m$ with $(i_1,i_2)=(2,1)$. This together with the previous round of phase $m$ implies the transmission of a total of $N^\text{X}_m$ order-$m$ symbols in $T^\text{X}_m$ time slots, and generation of $N^\text{X}_{m+1}$ order-$(m+1)$ symbols, where $N^\text{X}_m$, $T^\text{X}_m$, and $N^\text{X}_{m+1}$ are given by \cref{Eq:N'_m,Eq:T'_m,Eq:N'_m+1}.  If we deliver all these $\xxS_{m+1}$-symbols to their intended subsets of receivers, then each receiver will be able to decode all its desired order-$m$ symbols transmitted in this phase. This will be accomplished during the next phases.

\vspace{3mm}
$\bullet$ \underline{\textit{Phase $K$ ($2\times K$ X Channel)}:}
\vspace{3mm}

In this phase, during each time slot, an order-$K$ symbol of the form $u^{[i|\xxS_K]}$, $i\in \{1,2\}$, is transmitted by TX$_i$ while the other transmitter is silent. Therefore,
\begin{equation}
\label{Eq:DoF_K(K)-X1}
\DoFa^\text{X}_K(2,K)=1.
\end{equation}

Finally, using \cref{Eq:N'_m,Eq:T'_m,Eq:N'_m+1}, for any $1\leq m\leq K-1$, $\DoFa^\text{X}_m(2,K)$, the achieved DoF of transmission of order-$m$ symbols in the $2\times K$ SISO X channel with delayed CSIT, is given by
\begin{align}
\DoFa_m^\text{X}(2,K)&=\frac{N^\text{X}_m}{T^\text{X}_m+\frac{N^\text{X}_{m+1}}{\DoFa^\text{X}_{m+1}(2,K)}} \nonumber \\
&=\frac{2[2(K-m)+1]\binom{K}{m}}{2(K-m+1)\binom{K}{m}+\frac{2m\binom{K}{m+1}}{\DoFa^\text{X}_{m+1}(2,K)}} \nonumber \\
&= \frac{(m+1)[2(K-m)+1]}{(m+1)(K-m+1)+\frac{m(K-m)}{\DoFa^\text{X}_{m+1}(2,K)}}. \label{Eq:DoF_m^X(K)-X1}
\end{align}

It is proved in Appendix \ref{App:DoF-X} that \cref{Eq:DoF(K)-X,Eq:DoF_m(K)-X} are closed form expressions for $\DoFa_m^\text{X}(2,K)$, $1\leq m \leq K$, satisfying the recursive equation  \cref{Eq:DoF_m^X(K)-X1} together with the initial condition \eqref{Eq:DoF_K(K)-X1}.

\section{Conclusion}
\label{Sec:Conclusion}
We proposed multi-phase interference alignment schemes and obtained new achievable results on the DoF of the $K$-user SISO interference channel and $2\times K$ SISO X channel under delayed CSIT assumption. Our achievable DoFs are strictly greater than the best previously known DoFs for both channels with delayed CSIT and approach limiting values of $\frac{4}{6\ln 2 -1}$ and $\frac{1}{\ln 2}$, respectively, for the $K$-user interference and ${2\times K}$ X channel as $K\to \infty$. Without tight upper bounds, the problem of DoF characterization for both channels remains open. However, we conjecture that the DoF of both $K$-user interference and ${M\times K}$ X channel with delayed CSIT does not scale with the number of users.

\appendices

\section{Proof of Linear Independence in Phase $1$ for the $K$-user IC}
\label{App:LinearInependence}

In this appendix, we show that after phase $1$ of the proposed transmission scheme for the $K$-user SISO IC with delayed CSIT, the $(K-1)^2$ linear combinations obtained by each receiver in terms of its information symbols are linearly independent almost surely (see \cref{Sec:SISO-ICK}, \cref{Eq:Phase1-ICK1,Eq:Phase1-ICK2}). To this end, consider the aforementioned linear combinations at RX$_j$, $1\leq j \leq K$, \ie
\begin{align}
&(\xu^{[j]})^T \xQ_{jj}^T \xomega_{ji_1},\qquad i_1\in \xxS_K\backslash \{j\},\\
&(\xu^{[j]})^T \xQ_{i_2j}^T \xomega_{i_2i_3},\qquad \{i_2, i_3\}\subset \xxS_K\backslash \{j\},
\end{align}
which are equivalent to the system of linear combinations $(\xu^{[j]})^T\xP^{[j]}$, where $\xP^{[j]}$ is a $(K-1)^2\times (K-1)^2$ matrix defined as
\begin{align}
&\xP^{[j]} \Def \left[\left\{\xQ_{jj}^T \xomega_{ji_1} \right\}_{ i_1\in \xxS_K\backslash \{j\}}, \left\{ \xQ_{i_2j}^T \xomega_{i_2i_3}\right\}_{\{i_2, i_3\}\subset \xxS_K\backslash \{j\}}\right] \nonumber \\
&=(\xC^{[j]})^T \left[ \left\{ \xD_{jj}\xomega_{ji_1} \right\}_{ i_1\in \xxS_K\backslash \{j\}}, \left\{ \xD_{i_2j} \xomega_{i_2i_3}\right\}_{\{i_2, i_3\}\subset \xxS_K\backslash \{j\}} \right]. \nonumber
\end{align}
Let $\tilde{\xh}_{ij}$ denote the vector of length $(K-1)^2+1$ containing the main diagonal of $\xD_{ij}$ and define ${\xv_\ell \Def [\underbrace{1,1,\cdots,1}_\ell]^T}$. Then, one can write
\begin{equation}
\label{Eq:PMatrix_Expansion}
\xP^{[j]} = (\xC^{[j]})^T \left(\tilde{\xH}^{[j]} \circ \xOmega^{[j]} \right),
\end{equation}
where 
\begin{align}
\tilde{\xH}^{[j]} &\Def \left[\tilde{\xh}_{jj} \xv_{K-1}^T, \left\{ \tilde{\xh}_{ij} \xv_{K-2}^T \right\}_{i\in\xxS_{K}\backslash \{j\}}\right], \\
\xOmega^{[j]} &\Def \left[ \left\{\xomega_{ji_1} \right\}_{ i_1\in \xxS_K\backslash \{j\}}, \left\{ \xomega_{i_2i_3}\right\}_{\{i_2, i_3\}\subset \xxS_K\backslash \{j\}} \right] \nonumber \\
&= \left[\xomega_{j_1i_1} \right]_{i_1\in \xxS_K\backslash \{j\}, j_1\in \xxS_K \backslash \{i_1\}},
\end{align}
and ``$\circ$'' denotes the \emph{element-wise product} operator. Recall that $ \xQ_{j_1i_1}^T \xomega_{j_1i_1}=(\xC^{[i_1]})^T \xD_{j_1i_1}\xomega_{j_1i_1}=\mathbf{0}_{(K-1)^2\times 1}$. Hence, the vector $\xD_{j_1i_1}\xomega_{j_1i_1}$ lies in the left null space of $\xC^{[i_1]}$. However, $\xC^{[i_1]}$ is a random matrix of size $[(K-1)^2+1]\times (K-1)^2$, and thus, is full rank almost surely and its left null space is one dimensional, denoted by the nonzero unit vector $\xn^{[i_1]}$. It immediately follows that for any $j_1\in \xxS_K \backslash \{i_1\}$, there exists a nonzero scalar $a_{j_1i_1}$ such that $\xD_{j_1i_1}\xomega_{j_1i_1}=a_{j_1i_1}\xn^{[i_1]}$, or equivalently, $\xomega_{j_1i_1}=a_{j_1i_1}\xD^{-1}_{j_1i_1}\xn^{[i_1]}$. Note that $\xD_{j_1i_1}$ is full rank, and so, invertible almost surely. Therefore, $\xOmega^{[j]}$ can be rewritten as
\begin{align}
\xOmega^{[j]} =\left[a_{j_1i_1}\xD^{-1}_{j_1i_1}\xn^{[i_1]}\right]_{i_1\in \xxS_K\backslash \{j\}, j_1\in \xxS_K \backslash \{i_1\}}. \label{Eq:OmegaMatrix}
\end{align}
Since $a_{j_1i_1}$'s are nonzero and each of them scales a column of $\tilde{\xH}^{[j]} \circ \xOmega^{[j]}$, they do not affect the rank. Hence, 
\begin{align}
\rank&\left(\tilde{\xH}^{[j]}\circ \xOmega^{[j]}\right)= \rank\left(\tilde{\xH}^{[j]} \circ \left[\xD^{-1}_{j_1i_1}\xn^{[i_1]}\right]_{i_1\in \xxS_K\backslash \{j\}, j_1\in \xxS_K \backslash \{i_1\}}\right). \label{Eq:Rank1}
\end{align}
One also can write
\begin{align}
\tilde{\xH}^{[j]} \circ \left[\xD^{-1}_{j_1i_1}\xn^{[i_1]}\right]_{i_1\in \xxS_K\backslash \{j\}, j_1\in \xxS_K \backslash \{i_1\}} &= \tilde{\xH}^{[j]} \circ \xN^{[j]}\circ(\hat{\xH}^{[j]})^{\circ(-1)} \nonumber \\
&= \xPhi^{[j]} \circ (\hat{\xH}^{[j]})^{\circ(-1)}, \label{Eq:MatrixEq1}
\end{align}
where
\begin{align}
\hat{\xH}^{[j]} &\Def \left[ \tilde{\xh}_{j_1i_1}\right]_{i_1\in \xxS_K\backslash \{j\}, j_1\in \xxS_K \backslash \{i_1\}} \\
\xN^{[j]} &\Def \left[ \xn^{[i_1]} \xv_{K-1}^T\right]_{i_1\in \xxS_K\backslash \{j\}} \\
\xPhi^{[j]} &\Def  \tilde{\xH}^{[j]} \circ \xN^{[j]} ,
\end{align}
and $(\hat{\xH}^{[j]})^{\circ(-1)}$ denotes the element-wise inverse of $\hat{\xH}^{[j]}$. We note that $\hat{\xH}^{[j]}$ and $\xN^{[j]}$ are independent of each other, since $\xN^{[j]}$ is a function of $\{\xC^{[i_1]}\}_{i_1\in \xxS_K\backslash \{j\}}$ which are independent of $\hat{\xH}^{[j]}$. Also, $\tilde{\xH}^{[j]}$ and $\hat{\xH}^{[j]}$ are independent of each other, since the channel coefficients are i.i.d.\ across the transmitters and receivers. Hence, $\xPhi^{[j]}$ is independent of $\hat{\xH}^{[j]}$. 

On the other hand, it can be easily verified that the elements of $\hat{\xH}^{[j]}$, and thereby $(\hat{\xH}^{[j]})^{\circ(-1)}$, are i.i.d.. Also, it is easy to show that all elements of $\xPhi^{[j]}$ are nonzero almost surely. Therefore, for any given $\xPhi^{[j]}$, the elements of $\xPhi^{[j]} \circ (\hat{\xH}^{[j]})^{\circ(-1)}$ are also independent of each other, since $\xPhi^{[j]}$ is independent of $(\hat{\xH}^{[j]})^{\circ(-1)}$. This implies that for any given $\xPhi^{[j]}$, $\xPhi^{[j]} \circ (\hat{\xH}^{[j]})^{\circ(-1)}$ is full rank almost surely. This means that $\xPhi^{[j]} \circ (\hat{\xH}^{[j]})^{\circ(-1)}$ is full rank almost surely. 

Finally, we note that $\xC^{[j]}$ is independent of $\tilde{\xH}^{[j]}$, $\xN^{[j]}$, and $\hat{\xH}^{[j]}$, and thereby, of $\tilde{\xH}^{[j]} \circ \xOmega^{[j]}$. Therefore, regarding \cref{Eq:PMatrix_Expansion,Eq:Rank1,Eq:MatrixEq1} and applying \cref{Lm:Rank}, one can conclude that $\xP^{[j]}$ is full rank almost surely.

\begin{lemma}
\label{Lm:Rank}
Let $\xA_{m\times n}$ and $\xB_{n\times m}$ be two independent random matrices with continuous probability distributions and let $m\leq n$. If $\xA$ and $\xB$ are full rank almost surely, then $\vect{AB}$ is full rank almost surely.
\end{lemma}
\begin{IEEEproof}
If $m=n$, then the lemma is obviously true. Assume $m<n$. Let $\xa_i$, $1\leq i \leq n$, and $\vect{b}_j$, $1\leq j \leq m$, be the $i$'th and $j$'th column of $\xA$ and $\xB$, respectively. Then, the $j$'th column of $\vect{AB}$ can be written as $\sum_{i=1}^nb_{ji}\xa_i$. Now, assume a linear combination of the columns of $\vect{AB}$ is equal to zero, namely,
\begin{equation}
\sum_{j=1}^m\gamma_j\sum_{i=1}^nb_{ji}\xa_i = \vect{0}_{m\times 1}.
\end{equation}
Therefore, exchanging order of the summations, we have ${\sum_{i=1}^n\left(\sum_{j=1}^m\gamma_jb_{ji}\right)\xa_i=\vect{0}_{m\times 1}}$, which can be written in matrix form as
\begin{align}
\xA\sum_{j=1}^m\gamma_j\vect{b}_j = \vect{0}_{m\times 1}.
\end{align}
Thus, the vector $\sum_{j=1}^m\gamma_j\vect{b}_j$ either is equal to zero or lies in the null space of $\xA$. In the former case, we get $\gamma_j=0$, $1\leq j \leq m$, since $\xB$ is full rank almost surely. In the latter case, since $\xA$ is full rank almost surely, its null space is $n-m$ dimensional. Let $\xN_{n\times (n-m)}\Def \left[\xn_1, \xn_2, \cdots, \xn_{n-m}\right]$ denote a basis of the null space of $\xA$. Then, there should exist $\xi_\ell$, $1\leq \ell \leq n-m$, such that
\begin{equation}
\label{Eq:ANull}
\sum_{j=1}^m\gamma_j\vect{b}_j = \sum_{\ell=1}^{n-m}\xi_\ell \xn_\ell.
\end{equation}
Note that $\xN$ is independent of $\xB$, since $\xA$ and $\xB$ are independent of each other. Consider the square matrix $[\xB|\xN]_{n\times n}$. Since $\xB$ and $\xN$ are full rank almost surely (with continuous distributions) and independent of each other, one can easily show that $[\xB|\xN]$ is full rank almost surely. This together with \cref{Eq:ANull} yields $\gamma_j=0$, $1\leq j \leq m$, and $\xi_\ell=0$, $1\leq \ell \leq n-m$.
\end{IEEEproof}

\section{Proof of Linear Independence in Phase $m$-I for the $K$-user IC and Phase $m$ for the $K$-user X Channel}
\label{App:LinearInependence-S_m}
Consider the system of linear combinations
\begin{align}
&\xQ_{ji_1}\xu^{[i_1|\xxS_m]}+\xQ_{ji_2}\xu^{[i_2|\xxS_m]} \\
&(\xu^{[i_1|\xxS_m]})^T\xQ^T_{j'i_1}\xomega_{j'i_2}, \qquad j'\in \xxS_K \backslash \xxS_m,
\end{align}
which is equivalent to the system of linear combinations $(\xu^{[\xxS_m]})^T\xG^{[j]}$, where $\xG^{[j]}$ and $\xu^{[\xxS_m]}$ are defined as
\begin{align}
\xG^{[j]}&\Def \left[\begin{array}{c|c} (\xQ_{ji_1})^T &  \left\{\xQ^T_{j'i_1}\xomega_{j'i_2}\right\}_{j'\in \xxS_K \backslash \xxS_m}\\ \hline (\xQ_{ji_2})^T & \bigcirc \end{array}\right], \\
\xu^{[\xxS_m]} &\Def \left[(\xu^{[i_1|\xxS_m]})^T,(\xu^{[i_1|\xxS_m]})^T\right]^T.
\end{align}
Note first that by definition, $\xQ_{ji_1}=\xD_{ji_1}\xC^{[i_1|\xxS_m]}$ and $\xQ_{ji_2}=\xD_{ji_2}\xC^{[i_2|\xxS_m]}$. These matrix multiplications are nothing but scaling the columns of $\xC^{[i_1|\xxS_m]}$ and $\xC^{[i_2|\xxS_m]}$ by the diagonal elements of $\xD_{ji_1}$ and $\xD_{ji_2}$, respectively. Since the diagonal elements of $\xD_{ji_1}$ and $\xD_{ji_2}$ are nonzero almost surely and since scaling the columns of a matrix by nonzero factors does not affect its rank, one can write
\begin{align}
\rank\left(\xQ_{ji_1}\right)&=\rank\left(\xC^{[i_1|\xxS_m]}\right) = K-m+1,\\
\rank\left(\xQ_{ji_2}\right)&=\rank\left(\xC^{[i_2|\xxS_m]}\right) = K-m.
\end{align}

Also, if a linear combination of some columns is added to a (nonzero) scaled version of a column in a matrix then its rank does not change. Therefore, if we replace the $K-m+1$'th column of $\xG^{[j]}$ with a linear combination of its first $K-m+1$ columns, its rank will not change. If we choose the coefficients of such a linear combination to be the elements of $\xomega_{ji_2}$ (which are all nonzero almost surely), then since by definition, $ (\xQ_{ji_2})^T\xomega_{ji_2}=\vect{0}_{(K-m)\times 1}$, we get
\begin{equation}
\rank\left(\xG^{[j]}\right)=\rank\left(\tilde{\xG}^{[j]}\right),
\end{equation}
where
\begin{equation}
\tilde{\xG}^{[j]}\Def \left[\begin{array}{c|c} (\tilde{\xQ}_{ji_1})^T &  \left\{\xQ^T_{j'i_1}\xomega_{j'i_2}\right\}_{j'\in (\xxS_K \backslash \xxS_m)\cup \{j\}}\\ \hline (\tilde{\xQ}_{ji_2})^T & \bigcirc \end{array}\right], 
\end{equation}
and $\tilde{\xQ}_{ji_1}$ and $\tilde{\xQ}_{ji_2}$ are respectively the submatrices of $\xQ_{ji_1}$ and $\xQ_{ji_2}$ including their first $K-m$ rows. Hence, it suffices to show $\tilde{\xG}^{[j]}$ is full rank. To do so, we note that $\tilde{\xQ}_{ji_2}$ is a $(K-m)\times(K-m)$ matrix with $\rank(\tilde{\xQ}_{ji_2})=\rank(\xQ_{ji_2})=K-m$. If we show that the matrix $\left[\xQ^T_{j'i_1}\xomega_{j'i_2}\right]_{j'\in (\xxS_K \backslash \xxS_m)\cup \{j\}}$ is also a square full rank matrix of size $(K-m+1)\times(K-m+1)$, then using \cref{Lm:Rank2}, it immediately follows that  $\tilde{\xG}^{[j]}$ is full rank. Now, we rewrite $\left[\xQ^T_{j'i_1}\xomega_{j'i_2}\right]_{j'\in (\xxS_K \backslash \xxS_m)\cup \{j\}}$ as
\begin{align}
\left[\xQ^T_{j'i_1}\xomega_{j'i_2}\right]_{j'\in (\xxS_K \backslash \xxS_m)\cup \{j\}}=(\xC^{[i_1|\xxS_m]})^T\left[\xD_{j'i_1}\xomega_{j'i_2}\right]_{j'\in (\xxS_K \backslash \xxS_m)\cup \{j\}}.
\end{align}
Since the matrices are square, we have 
\begin{align}
\det\left(\left[\xQ^T_{j'i_1}\xomega_{j'i_2}\right]_{j'\in (\xxS_K \backslash \xxS_m)\cup \{j\}}\right) =\det\left(\xC^{[i_1|\xxS_m]}\right) \cdot \det \left(\left[\xD_{j'i_1}\xomega_{j'i_2}\right]_{j'\in (\xxS_K \backslash \xxS_m)\cup \{j\}}\right), \nonumber
\end{align}
and since $\xC^{[i_1|\xxS_m]}$ is full rank almost surely, $\det\left(\xC^{[i_1|\xxS_m]}\right)\neq0$. Thus, it remains to show $\left[\xD_{j'i_1}\xomega_{j'i_2}\right]_{j'\in (\xxS_K \backslash \xxS_m)\cup \{j\}}$ is full rank. Using the same argument as in Appendix \ref{App:LinearInependence}, one can write
\begin{equation}
\xomega_{j'i_2}=a_{j'i_2}\xD^{-1}_{j'i_2}\xn^{[i_2]},
\end{equation}
where $a_{j'i_2}$ is a nonzero scalar. Therefore, 
\begin{align}
\rank \left(\left[\xD_{j'i_1}\xomega_{j'i_2}\right]_{j'\in (\xxS_K \backslash \xxS_m)\cup \{j\}}\right)&=\rank\left(\left[a_{j'i_2}\xD_{j'i_1}\xD^{-1}_{j'i_2}\xn^{[i_2]}\right]_{j'\in (\xxS_K \backslash \xxS_m)\cup \{j\}}\right) \nonumber \\
&\stackrel{\text{(a)}}{=}\rank\left(\left[\xD_{j'i_1}\xD^{-1}_{j'i_2}\xn^{[i_2]}\right]_{j'\in (\xxS_K \backslash \xxS_m)\cup \{j\}}\right) \nonumber \\
&\stackrel{\text{(b)}}{=}\rank\bigg(\left[\frac{h_{j'i_1}(t)}{h_{j'i_2}(t)}\right]_{\substack{1\leq t \leq K-m+1 \\j'\in (\xxS_K \backslash \xxS_m)\cup \{j\}}}\bigg) \nonumber \\
&\stackrel{\text{(c)}}{=}K-m+1, 
\end{align}
where (a) follows from the fact that scaling the columns of a matrix by nonzero factors ($a_{j'i_2}$'s) will not change its rank; (b) follows from the fact that scaling the rows of a matrix by nonzero factors (elements of $\xn^{[i_2]}$) will not change its rank; and (c) is true since $\frac{h_{j'i_1}(t)}{h_{j'i_2}(t)}$'s are i.i.d. for $1\leq t \leq K-m+1$ and $j'\in (\xxS_K \backslash \xxS_m)\cup \{j\}$.
\begin{lemma}
\label{Lm:Rank2}
Let $\xA=\left[a_{ij}\right]_{m\times m}$ and $\xB=\left[b_{ij}\right]_{n\times n}$ be two square matrices which are full rank almost surely and let $\xC=\left[c_{ij}\right]_{m\times n}$ be an arbitrary matrix. Then, the matrix
\begin{align}
\xD= \left[\begin{array}{c|c} \xC& \xA \\ \hline \xB & \bigcirc \end{array}\right] \nonumber
\end{align}
is full rank almost surely.
\end{lemma}
\begin{IEEEproof}
Denote by $\xa_j$, $\xb_j$, and $\xd_j$ the $j$'th columns of $\xA$, $\xB$, and $\xD$, respectively. Assume that 
\begin{equation}
\label{Eq:ZeroLinearCombination}
\sum_{j=1}^{m+n}\alpha_j\xd_j=\mathbf{0}_{(m+n) \times 1},
\end{equation}
for some $\alpha_1, \alpha_2, \cdots, \alpha_{m+n}\in \mathbb{C}$. Then, since $d_{ij}=0$ for $m+1\leq i \leq m+n$ and $n+1 \leq j \leq m+n$, one can write $\sum_{j=1}^{n}\alpha_j\xb_j=\mathbf{0}_{n \times 1}$, and since $\xB$ is full rank almost surely, we have $\alpha_j=0$, $1\leq j \leq n$. This together with \cref{Eq:ZeroLinearCombination} yields $\sum_{j=n+1}^{m+n}\alpha_j\xd_j=\mathbf{0}_{(m+n) \times 1}$. Considering the first $m$ elements of these columns, it follows that $\sum_{j=n+1}^{m+n}\alpha_j\xa_{j-n}=\mathbf{0}_{m \times 1}$, and since $\xA$ is full rank almost surely, we have $\alpha_j=0$, $n+1\leq j \leq m+n$.
\end{IEEEproof}

\section{Closed Form Solution to the Recursive Equation \eqref{Eq:DoF_m(K)-IC1} for the $K$-user IC}
\label{App:DoF-IC}
In this appendix, we derive a closed form solution to the recursive equation
\begin{align}
\DoFa_{K-i}^\text{IC}(K)&=\frac{(K-i)(2i+1)}{(K-i)(i+1)+\frac{i}{K-i+1}+\frac{(K-i-1)i}{\DoFa_{K-i+1}^\text{IC}(K)}}, \nonumber \\
&\hspace{30mm} 1\leq i\leq K-2, \label{Eq:IC-Rec1}\\
\DoFa_K^\text{IC}(K)&=1. \label{Eq:IC-Rec2}
\end{align}

We start by defining $A_{K-i}(K)\Def1-\frac{1}{\DoFa^\text{IC}_{K-i}(K)}$. Then, for $1\leq i \leq K-2$, we have
\begin{align}
A_{K-i}(K)=\frac{i}{(K-i)(2i+1)}\times \left[(K-i-1)A_{K-i+1}(K)+\frac{K-i}{K-i+1}\right], \label{Eq:A_K-i}
\end{align}
with $A_K(K)=0$. Express $A_{K-i}(K)$ as
\begin{align}
A_{K-i}(K)=\sum_{\ell=0}^{i}\frac{a_{K-\ell}^{[K-i]}}{K-\ell}, \label{Eq:A_K-i-sum_1}
\end{align}
where $a_{K-\ell}^{[K-i]}$ is given by
\begin{equation}
\label{Eq:a_K-l}
a_{K-\ell}^{[K-i]} = \left[(K-\ell)A_{K-i}(K)\right]\big|_{K=\ell}, \qquad 0\leq \ell \leq i.
\end{equation}
Substituting the expansion of \cref{Eq:A_K-i-sum_1} for $A_{K-i+1}(K)$ in \cref{Eq:A_K-i}, we get
\begin{align}
A_{K-i}(K)= \frac{i}{(K-i)(2i+1)}\times\left[\sum_{\ell=0}^{i-1}\frac{(K-i-1)a_{K-\ell}^{[K-i+1]}}{K-\ell}+\frac{K-i}{K-i+1}\right]. \label{Eq:A_K-i_1}
\end{align}
\Cref{Eq:a_K-l,Eq:A_K-i_1} lead to three recursive equations
\begin{align}
a_{K-\ell}^{[K-i]}&=\frac{(i-\ell+1)i}{(i-\ell)(2i+1)}a_{K-\ell}^{[K-i+1]}, \hspace{3mm} 0\leq \ell \leq i-2, \label{Eq:a_K-l_1} \\
a_{K-i+1}^{[K-i]}&=\frac{i}{2i+1}\left( 2a_{K-i+1}^{[K-i+1]}+1\right), \label{Eq:a_K-i+1}\\
a_{K-i}^{[K-i]}&=-\frac{i}{2i+1}\sum_{\ell=0}^{i-1}\frac{a_{K-\ell}^{[K-i+1]}}{i-\ell} \nonumber \\
&= -\frac{i}{2i+1}a_{K-i+1}^{[K-i+1]}-\sum_{\ell=0}^{i-2}\frac{a_{K-\ell}^{[K-i]}}{i-\ell+1}, \label{Eq:a_K-i}
\end{align}
where \cref{Eq:a_K-i} follows from \cref{Eq:a_K-l_1}. Applying \cref{Eq:a_K-l_1} $i-\ell-1$ times, we will have
\begin{align}
a_{K-\ell}^{[K-i]}&=\frac{1}{2}a_{K-\ell}^{[K-\ell-1]}(i-\ell+1)\prod_{j=\ell+2}^i\frac{j}{2j+1}, \label{Eq:a_K-l_2}
\end{align}
for any $0\leq \ell \leq i-2$. Substituting \cref{Eq:a_K-l_2} in \cref{Eq:a_K-i}, we get
\begin{align}
a_{K-i}^{[K-i]}&\stackrel{\text{(a)}}{=} -\frac{i}{2i+1}a_{K-i+1}^{[K-i+1]}-\frac{1}{2}\sum_{\ell=0}^{i-2}a_{K-\ell}^{[K-\ell-1]}\prod_{j=\ell+2}^i\frac{j}{2j+1} \nonumber \\
&\stackrel{\text{(b)}}{=} -\frac{i}{2i+1}\bigg[-\frac{i-1}{2(i-1)+1)}a_{K-i+2}^{[K-i+2]}-\frac{1}{2}\sum_{\ell=0}^{i-3}a_{K-\ell}^{K-\ell-1}\prod_{j=\ell+2}^{i-1}\frac{j}{2j+1} \bigg]-\frac{1}{2}\sum_{\ell=0}^{i-2}a_{K-\ell}^{[K-\ell-1]}\prod_{j=\ell+2}^i\frac{j}{2j+1} \nonumber \\
&=\frac{i(i-1)a_{K-i+2}^{[K-i+2]}}{(2i+1)\left[2(i-1)+1\right]}-\frac{i}{2(2i+1)}a_{K-i+2}^{[K-i+1]}\nonumber \\
&\stackrel{\text{(c)}}{=}\frac{i(i-1)}{(2i+1)\left[2(i-1)+1\right]}a_{K-i+2}^{[K-i+2]}-\frac{i}{2(2i+1)}\times \frac{i-1}{2(i-1)+1}\left( 2a_{K-i+2}^{[K-i+2]}+1\right) \nonumber \\
&=-\frac{i(i-1)}{2(2i+1)\left[ 2(i-1)+1\right]} \nonumber \\
&=-\frac{i(i-1)}{2(4i^2-1)}, \hspace{15mm}0\leq i \leq K-2, \label{Eq:a_K-i_2}
\end{align}
where (b) results from reapplying (a) to $a_{K-i+1}^{[K-i+1]}$, and (c) follows from applying \cref{Eq:a_K-i+1} to $a_{K-i+2}^{K-i+1}$.

Employing \cref{Eq:a_K-i_2} for $a_{K-i+1}^{[K-i+1]}$ in \cref{Eq:a_K-i+1}, one obtains
\begin{align}
a_{K-i+1}^{[K-i]}&=\frac{i}{2i+1}\left[1-\frac{(i-1)(i-2)}{4(i-1)^2-1}\right]\nonumber \\
 &=\frac{i}{2i+1}\times \frac{3(i-1)^2+(i-1)-1}{4(i-1)^2-1}, \label{Eq:a_K-i+1_2}
\end{align}
for any $0\leq i \leq K-2$. It follows from plugging \cref{Eq:a_K-i+1_2} into \cref{Eq:a_K-l_2} that for any $0\leq \ell \leq i-2$,
\begin{align}
a_{K-\ell}^{[K-i]}=\frac{(i-\ell+1)(3\ell^2+\ell-1)}{2(4\ell^2-1)}\prod_{j=\ell+1}^i\frac{j}{2j+1}. \label{Eq:a_K-l_3}
\end{align}
Finally, using \cref{Eq:A_K-i-sum_1,Eq:a_K-i_2,Eq:a_K-i+1_2,Eq:a_K-l_3}, we have
\begin{align}
A_{K-i}(K)=-&\frac{i(i-1)}{2(4i^2-1)(K-i)}+\sum_{\ell=0}^{i-1}\frac{(i-\ell+1)(3\ell^2+\ell-1)}{2(K-\ell)(4\ell^2-1)}\prod_{j=\ell+1}^{i}\frac{j}{2j+1}, \nonumber
\end{align}
for any $0\leq i \leq K-2$. Since by definition, $\DoFa_{K-i}^\text{IC}(K)= \frac{1}{1-A_{K-i}(K)}$, we have the following closed form expression for $\DoFa_{K-i}^\text{IC}(K)$, $0\leq i \leq K-2$,
\begin{align}
\DoFa_{K-i}^\text{IC}&(K)=\bigg[1+\frac{i(i-1)}{2(4i^2-1)(K-i)}-\sum_{\ell=0}^{i-1}\frac{(i-\ell+1)(3\ell^2+\ell-1)}{2(K-\ell)(4\ell^2-1)}\prod_{j=\ell+1}^{i}\frac{j}{2j+1}\bigg]^{-1}.
\end{align}

\section{Closed Form Solution to the Recursive Equation \eqref{Eq:DoF_m^X(K)-X1} for the $2\times K$ X Channel}
\label{App:DoF-X}
In this appendix, we derive the closed form solution to the recursive equation
\begin{align}
\DoFa_{K-i}^\text{X}(2,K)&= \frac{(K-i+1)(2i+1)}{(K-i+1)(i+1)+\frac{(K-i)i}{\DoFa^\text{X}_{K-i+1}(2,K)}}, \nonumber \\
&\hspace{30mm}1\leq i\leq K-1, \label{Eq:X-Rec1} \\
\DoFa^\text{X}_K(2,K)&=1. \label{Eq:X-Rec2}
\end{align}

Defining $B_{K-i}(K)\Def1-\frac{1}{\DoFa_{K-i}^\text{X}(2,K)}$, for any $1\leq i \leq K-1$, one can write
\begin{align}
B_{K-i}(K)&=\frac{i\times \left[(K-i)B_{K-i+1}(K)+1\right]}{(K-i+1)(2i+1)}, \label{Eq:B_K-i}
\end{align}
with $B_K(K)=0$. Express $B_{K-i}(K)$ as
\begin{equation}
\label{Eq:B_K-i-sum_1}
B_{K-i}(K)=\sum_{\ell=0}^{i-1}\frac{b_{K-\ell}^{[K-i]}}{K-\ell},
\end{equation}
where $b_{K-\ell}^{[K-i]}$ is given by
\begin{equation}
\label{Eq:b_K-l}
b_{K-\ell}^{[K-i]} = \left[(K-\ell)B_{K-i}(K)\right]\big|_{K=\ell},\hspace{5mm} 0\leq \ell \leq i-1.
\end{equation}
Substituting the expansion of \cref{Eq:B_K-i-sum_1} for $B_{K-i+1}(K)$ in \cref{Eq:B_K-i}, we get
\begin{align}
B_{K-i}(K)=\frac{i}{(K-i+1)(2i+1)}\left[\sum_{\ell=0}^{i-2}\frac{(K-i)b_{K-\ell}^{[K-i+1]}}{K-\ell}+1\right]. \label{Eq:B_K-i_1}
\end{align}

\Cref{Eq:b_K-l,Eq:B_K-i_1} result in two recursive equations
\begin{align}
b_{K-\ell}^{[K-i]}&=\frac{i(i-\ell)b_{K-\ell}^{[K-i+1]}}{(i-\ell-1)(2i+1)}, \hspace{5mm} 0\leq \ell \leq i-2, \label{Eq:b_K-l_1} \\
b_{K-i+1}^{[K-i]}&=\frac{i}{2i+1}\left[1-\sum_{\ell=0}^{i-2}\frac{b_{K-\ell}^{[K-i+1]}}{i-\ell-1} \right]\nonumber \\
&=\frac{i}{2i+1}-\sum_{\ell=0}^{i-2}\frac{b_{K-\ell}^{[K-i]}}{i-\ell}, \label{Eq:b_K-i+1}
\end{align}
where \cref{Eq:b_K-i+1} follows from \cref{Eq:b_K-l_1}. Applying \cref{Eq:b_K-l_1} $i-\ell-1$ times, we will have
\begin{align}
b_{K-\ell}^{[K-i]}&=b_{K-\ell}^{[K-\ell-1]}(i-\ell)\prod_{j=\ell+2}^i\frac{j}{2j+1}, \label{Eq:b_K-l_2}
\end{align} 
for any $0\leq \ell \leq i-2$. Substituting \cref{Eq:b_K-l_2} in \cref{Eq:b_K-i+1}, it follows that
\begin{align}
b_{K-i+1}^{[K-i]}&\stackrel{\text{(a)}}{=}\frac{i}{2i+1}-\sum_{\ell=0}^{i-2}b_{K-\ell}^{[K-\ell-1]}\prod_{j=\ell+2}^i\frac{j}{2j+1} \nonumber \\
&=\frac{i}{2i+1}-\frac{i}{2i+1}b_{K-i+2}^{[K-i+1]}-\sum_{\ell=0}^{i-3}b_{K-\ell}^{[K-\ell-1]}\prod_{j=\ell+2}^i\frac{j}{2j+1} \nonumber \\
&=\frac{i}{2i+1}-\frac{i}{2i+1}b_{K-i+2}^{[K-i+1]}-\frac{i}{2i+1}\sum_{\ell=0}^{i-3}b_{K-\ell}^{[K-\ell-1]}\prod_{j=\ell+2}^{i-1}\frac{j}{2j+1} \nonumber \\
&=\frac{i}{2i+1}-\frac{i}{2i+1}\times \bigg[b_{K-i+2}^{[K-i+1]}+\sum_{\ell=0}^{i-3}b_{K-\ell}^{[K-\ell-1]}\prod_{j=\ell+2}^{i-1}\frac{j}{2j+1}\bigg] \nonumber \\
&\stackrel{\text{(b)}}{=}\frac{i}{2i+1}-\frac{i}{2i+1}\times\frac{i-1}{2(i-1)+1} \nonumber \\
&=\frac{i^2}{4i^2-1},\hspace{15mm} 0\leq i \leq K-1, \label{Eq:b_K-i+1_2}
\end{align}
where (b) simply follows from an application of (a) for $b_{K-i+2}^{[K-i+1]}$. Substituting \cref{Eq:b_K-i+1_2} for $b_{K-\ell}^{[K-\ell-1]}$ in \cref{Eq:b_K-l_2}, we obtain
\begin{equation}
\label{Eq:b_K-l_3}
b_{K-\ell}^{[K-i]}=\frac{(i-\ell)(\ell+1)}{2(\ell+1)-1}\prod_{j=\ell+1}^i\frac{j}{2j+1}, \hspace{2mm}0\leq \ell \leq i-2.
\end{equation}
Combining \cref{Eq:B_K-i-sum_1,Eq:b_K-i+1_2,Eq:b_K-l_3}, we can write
\begin{align}
B_{K-i}(K)&=\sum_{\ell=0}^{i-1}\frac{(i-\ell)(\ell+1)}{(K-\ell)\left[2(\ell+1)-1\right]}\prod_{j=\ell+1}^i\frac{j}{2j+1}, \nonumber
\end{align}
which together with $\DoFa_{K-i}^\text{X}(2,K)= \frac{1}{1-B_{K-i}(K)}$ yields
\begin{align}
\DoFa_{K-i}^\text{X}(2,K)= \bigg[ 1-\sum_{\ell=0}^{i-1}\frac{(i-\ell)(\ell+1)}{(K-\ell)(2\ell+1)}\prod_{j=\ell+1}^{i}\frac{j}{2j+1}\bigg]^{-1}, \nonumber
\end{align}
for any $0\leq i \leq K-1$.

\section{Asymptotic Behavior of the Achievable DoFs}
\label{App:DoF-Limit}
In this appendix, we show that
\begin{align}
\lim_{K\to \infty} \DoFa_1^\text{IC}(K)&=\frac{4}{6\ln 2 -1}, \\
\lim_{K\to \infty} \DoFa_1^\text{X}(2,K)&=\frac{1}{\ln2}.
\end{align}

In view of \cref{Eq:DoF(K)-IC,Eq:A_2(K),Eq:DoF(K)-X}, it suffices to show that
\begin{align}
\lim_{K\to \infty} \Psi(K) &=\frac{21}{16}-\frac{3}{2}\ln 2. \label{Eq:Psi}\\
\lim_{K\to \infty} \Phi(K) &=1-\ln 2. \label{Eq:Phi}
\end{align}
where
\begin{align}
\Psi(K) &\Def \sum_{\ell_1=0}^{K-3}\frac{(K-\ell_1-1)(3\ell_1^2+\ell_1-1)}{2(K-\ell_1)(4\ell_1^2-1)}\prod_{\ell_2=\ell_1+1}^{K-2}\frac{\ell_2}{2\ell_2+1}, \nonumber \\
\Phi(K) &\Def \sum_{\ell_1=0}^{K-2}\frac{(K-\ell_1-1)(\ell_1+1)}{(K-\ell_1)(2\ell_1+1)}\prod_{\ell_2=\ell_1+1}^{K-1}\frac{\ell_2}{2\ell_2+1}. \nonumber
\end{align}
To do so, for integers $K,p\geq0$, define $\Gamma_p(K)$ and $\Lambda_p(K)$ as
\begin{align}
\Gamma_p(K)&\Def\sum_{\ell=0}^{K-p}\frac{K-\ell-1}{(K-\ell)2^{K-\ell}}, \\
\Lambda_p(K)&\Def\sum_{\ell=0}^{K-p}\frac{\ell(K-\ell-1)}{K(K-\ell)2^{K-\ell}}.
\end{align}
Using $\sum_{n=1}^\infty \frac{1}{n2^n}=\ln2$, $\sum_{n=1}^\infty \frac{n}{2^n}=2$, and $\sum_{n=1}^\infty \frac{1}{2^n}=1$, it is easily verified that, for any integer $p\geq0$,
\begin{align}
\lim_{K\to \infty}\Gamma_p(K)=\lim_{K\to \infty}\Lambda_p(K)=-\ln2+2^{1-p}+\sum_{n=1}^{p-1}\frac{1}{n2^n}. \nonumber
\end{align}
Specifically, 
\begin{align}
\lim_{K\to \infty}\Gamma_2(K)=\lim_{K\to \infty}\Lambda_2(K)=1-\ln 2, \\
\lim_{K\to \infty}\Gamma_3(K)=\lim_{K\to \infty}\Lambda_3(K)=\frac{7}{8}-\ln 2.
\end{align}
Now, using the following two lemmas together with the Squeeze Theorem, \cref{Eq:Psi,Eq:Phi} are immediate.
\begin{lemma}
The following inequalities hold for $K\geq 3$.
\begin{align}
\frac{3K}{2K-3} \Lambda_3(K)<\Psi(K)< \frac{3}{2}\Gamma_3(K)+\frac{K-2}{5(K-1)2^K}. \nonumber
\end{align}
\end{lemma}
\begin{IEEEproof}
\begin{itemize}
\item[(i)] Upper bound:
\end{itemize}
\begin{align}
\Psi(K)&= \sum_{\ell_1=0}^{K-3}\frac{(K-\ell_1-1)(3\ell_1^2+\ell_1-1)}{2(K-\ell_1)(4\ell_1^2-1)}\prod_{\ell_2=\ell_1+1}^{K-2}\frac{\ell_2}{2\ell_2+1} \nonumber \\
& =\sum_{\ell_1=0}^{K-3}\frac{(K-\ell_1-1)(3\ell_1^2+\ell_1-1)(\ell_1+1)}{2(K-\ell_1)(4\ell_1^2-1)(2\ell_1+3)}\prod_{\ell_2=\ell_1+2}^{K-2}\frac{\ell_2}{2\ell_2+1} \nonumber\\
& =\frac{K-1}{6K}\prod_{\ell_2=2}^{K-2}\frac{\ell_2}{2\ell_2+1}+\frac{K-2}{5(K-1)}\prod_{\ell_2=3}^{K-2}\frac{\ell_2}{2\ell_2+1}+\sum_{\ell_1=2}^{K-3}\frac{(K-\ell_1-1)(3\ell_1^2+\ell_1-1)(\ell_1+1)}{2(K-\ell_1)(4\ell_1^2-1)(2\ell_1+3)}\prod_{\ell_2=\ell_1+2}^{K-2}\frac{\ell_2}{2\ell_2+1}\nonumber\\
& \stackrel{\text{(a)}}{<} \frac{K-2}{5\times2^4(K-1)}\prod_{\ell_2=3}^{K-2}\frac{\ell_2}{2\ell_2+1}+\bigg\{\frac{3(K-1)}{2^4K}\prod_{\ell_2=2}^{K-2}\frac{\ell_2}{2\ell_2+1}{+}\frac{3(K-2)}{2^4(K-1)}\prod_{\ell_2=3}^{K-2}\frac{\ell_2}{2\ell_2+1}\nonumber\\
&\hspace{26mm} +\sum_{\ell_1=2}^{K-3}\frac{3(K-\ell_1-1)}{2^4(K-\ell_1)}\prod_{\ell_2=\ell_1+2}^{K-2}\frac{\ell_2}{2\ell_2+1}\bigg\} \nonumber\\
& =\frac{K-2}{5\times2^4(K-1)}\prod_{\ell_2=3}^{K-2}\frac{\ell_2}{2\ell_2+1}+\sum_{\ell_1=0}^{K-3}\frac{3(K-\ell_1-1)}{2^4(K-\ell_1)}\prod_{\ell_2=\ell_1+2}^{K-2}\frac{\ell_2}{2\ell_2+1}\nonumber\\
& \stackrel{\text{(b)}}{<} \frac{K-2}{5(K-1)2^K}+\frac{3}{2}\sum_{\ell_1=0}^{K-3}\frac{K-\ell_1-1}{(K-\ell_1)2^{K-\ell_1}} \nonumber\\
& = \frac{3}{2} \Gamma_3(K)+\frac{K-2}{5(K-1)2^K},
\end{align}
where (a) follows from the fact that $\frac{(3\ell_1^2+\ell_1-1)(\ell_1+1)}{(4\ell_1^2-1)(2\ell_1+3)}<\frac{3}{8}$ for $\ell_1\geq2$ together with inequality $\frac{1}{6}<\frac{3}{16}$, and (b) is valid since $\frac{\ell_2}{2\ell_2+1}<\frac{1}{2}$ for $\ell_2\geq2$.
\begin{itemize}
\item[(ii)] Lower bound:
\end{itemize}
\begin{align} 
\Psi(K)& =\sum_{\ell_1=0}^{K-3}\frac{(K-\ell_1-1)(3\ell_1^2+\ell_1-1)}{2(K-\ell_1)(4\ell_1^2-1)}\prod_{\ell_2=\ell_1+1}^{K-2}\frac{\ell_2}{2\ell_2+1} \nonumber\\
&= \sum_{\ell_1=0}^{K-3}\frac{(K-\ell_1-1)(3\ell_1^2+\ell_1-1)(\ell_1+1)}{2(2K-3)(K-\ell_1)(4\ell_1^2-1)}\prod_{\ell_2=\ell_1+2}^{K-2}\frac{\ell_2}{2\ell_2-1} \nonumber\\
&\stackrel{\text{(a)}}{>} \frac{3K}{2K-3}\sum_{\ell_1=0}^{K-3}\frac{\ell_1(K-\ell_1-1)}{K(K-\ell_1)2^{K-\ell_1}} \nonumber\\
&= \frac{3K}{2K-3} \Lambda_3(K),
\end{align}
where (a) follows from the fact that $\frac{(3\ell_1^2+\ell_1-1)(\ell_1+1)}{4\ell_1^2-1}>\frac{3}{4}\ell_1$ for $\ell_1\geq 0$, and $\frac{\ell_2}{2\ell_2-1}>\frac{1}{2}$ for $\ell_2\geq2$.
\end{IEEEproof}

\begin{lemma}
The following inequalities hold for $K\geq2$.
\begin{align}
\frac{2K\Lambda_2(K)}{2K{-}1}<\Phi(K)< \Gamma_3(K)+\frac{(K{-}1)^2}{2(2K{-}1)(2K{-}3)}+\frac{K{-}1}{15K2^K}. \nonumber
\end{align}
\end{lemma}
\begin{IEEEproof}
\begin{itemize}
\item[(i)] Upper bound:
\end{itemize}
\begin{align}
\Phi(K)&=\sum_{\ell_1=0}^{K-2}\frac{(K-\ell_1-1)(\ell_1+1)}{(K-\ell_1)(2\ell_1+1)}\prod_{\ell_2=\ell_1+1}^{K-1}\frac{\ell_2}{2\ell_2+1} \nonumber\\
&=\frac{(K-1)^2}{2(2K-1)(2K-3)}+\sum_{\ell_1=0}^{K-3}\frac{(K-\ell_1-1)(\ell_1+1)^2(\ell_1+2)}{(K-\ell_1)(2\ell_1+1)(2\ell_1+3)(2\ell_1+5)}\prod_{\ell_2=\ell_1+3}^{K-1}\frac{\ell_2}{2\ell_2+1} \nonumber\\
&\stackrel{\text{(a)}}{<}\frac{(K-1)^2}{2(2K-1)(2K-3)}+\frac{2(K-1)}{15K}\prod_{\ell_2=3}^{K-1}\frac{\ell_2}{2\ell_2+1}+\sum_{\ell_1=1}^{K-3}\frac{K-\ell_1-1}{2^3(K-\ell_1)}\prod_{\ell_2=\ell_1+3}^{K-1}\frac{\ell_2}{2\ell_2+1} \nonumber \\
&=\frac{(K-1)^2}{2(2K-1)(2K-3)}+\frac{(K-1)}{15\times2^3K}\prod_{\ell_2=3}^{K-1}\frac{\ell_2}{2\ell_2+1}+\sum_{\ell_1=0}^{K-3}\frac{K-\ell_1-1}{2^3(K-\ell_1)}\prod_{\ell_2=\ell_1+3}^{K-1}\frac{\ell_2}{2\ell_2+1} \nonumber \\
&\stackrel{\text{(b)}}{<}\frac{(K-1)^2}{2(2K-1)(2K-3)}+\frac{(K-1)}{15K2^K}+\sum_{\ell_1=0}^{K-3}\frac{K-\ell_1-1}{(K-\ell_1)2^{K-\ell_1}} \nonumber \\
&=\Gamma_3(K)+\frac{(K-1)^2}{2(2K-1)(2K-3)}+\frac{K-1}{15K2^K},
\end{align}
where (a) follows from the fact that $\frac{(\ell_1+1)^2(\ell_1+2)}{(2\ell_1+1)(2\ell_1+3)(2\ell_1+5)}<\frac{1}{8}$ for $\ell_1\geq 1$, and (b) is true since $\frac{\ell_2}{2\ell_2+1}<\frac{1}{2}$ for $\ell_2\geq3$.
\begin{itemize}
\item[(ii)] Lower bound:
\end{itemize}
\begin{align}
\Phi(K)&=\sum_{\ell_1=0}^{K-2}\frac{(K-\ell_1-1)(\ell_1+1)}{(K-\ell_1)(2\ell_1+1)}\prod_{\ell_2=\ell_1+1}^{K-1}\frac{\ell_2}{2\ell_2+1} \nonumber\\
&= \sum_{\ell_1=0}^{K-2}\frac{(K-\ell_1-1)(\ell_1+1)^2}{(2K-1)(K-\ell_1)(2\ell_1+1)}\prod_{\ell_2=\ell_1+2}^{K-1}\frac{\ell_2}{2\ell_2-1} \nonumber\\
&\stackrel{\text{(a)}}{>} \frac{2K}{2K-1}\sum_{\ell_1=0}^{K-2}\frac{\ell_1(K-\ell_1-1)}{K(K-\ell_1)2^{K-\ell_1}} \nonumber\\
&=\frac{2K}{2K-1} \Lambda_2(K),
\end{align}
where (a) follows from the fact that $\frac{(\ell_1+1)^2}{2\ell_1+1}>\frac{1}{2}\ell_1$ for $\ell_1\geq0$, and $\frac{\ell_2}{2\ell_2-1}>\frac{1}{2}$ for $\ell_2\geq2$.
\end{IEEEproof}

\bibliographystyle{IEEEtran}
\bibliography{SISO_IC_X_DelayedCSIT_FinalVersion}
\end{document}